\documentclass[a4paper,11pt]{article}
\pdfoutput=1 
             
\usepackage{jcappub} 

\newcommand{\diff}{\text{Diff}}

\newcommand{\gl}{GL(\mathbb{R}^4)}
\renewcommand{\b}{\overline}
\newcommand{\R}{\mathbb{R}}
\newcommand{\D}{\mathcal{D}}
\newcommand{\F}{\mathcal{F}}
\newcommand{\n}{\boldsymbol{n}}
\renewcommand{\u}{\boldsymbol{u}}
\renewcommand{\S}{\mathrm{S}}
\renewcommand{\t}{\mathrm{t}}

\usepackage{bbm} 
\usepackage{subcaption} 

\title{\boldmath Cosmic averaging over \\ multiscaled structure:\\ on foliations, gauges and backreaction}

\author{Dave B. H. Verweg,}
\author{Bernard J. T. Jones}
\author{and Rien van de Weygaert}
\affiliation{Kapteyn Astronomical Institute, University of Groningen, The Netherlands}
\emailAdd{dave@astro.verweg.net}

\abstract{
    The observation that accelerated cosmic expansion is dominant since the Megaparsec cosmic structure became nonlinear seems like an extraordinary coincidence, unless the acceleration is somehow driven by the emergence of the structure. That has given rise to the controversial concept of a gravitational backreaction through which inhomogeneity becomes a driver of accelerated expansion. 
    
    The standard route when studying strongly inhomogeneous cosmological models is to take either a perturbative approach or a spatial averaging approach. Here we argue that because backreaction is in fact a nonlinear \textit{multiscale} phenomenon, perturbative approaches may have a limited validity. The alternative is the proposed averaging approach. In  this paper we demonstrate that the implied backreaction terms are artificial, that is gauge dependent, which may easily cause ambiguous estimates of its significance.

    In the current study, we forward a formal fully geometric framework of cosmic foliations in the context of relativistic cosmology. Here we show that fixing a foliation of spacetime determines a choice of gauge. Addressing the correspondence between the metric tensor and the foliation allows us to clarify the theoretical implications of choosing a foliation.
    
    Within the context of backreaction, this formalism allows us to discuss the complications of averaging. It reveals that spatial averaging can induce artificial backreaction terms that arise from any specific choice of gauge. Averaging methods presented so far all encounter this problem. Within our foliation framework, we can produce a gauge invariant method of averaging by considering a group of proper time foliations which any cosmic observe can agree upon. We demonstrate that this implies the gauge invariance of the averaging procedure. This makes it applicable to standard cosmological simulations. 
}

\begin{document}
\maketitle
\flushbottom


\section{Introduction}
\label{sec:intro}
Over two decades ago, the Supernova Cosmology Project \cite{perlmutter1998discovery} and the High-Z Supernova Search Team \cite{riess1998observational} found that the expansion of the Universe is accelerating.  This suggested the existence of a dominant 74\% contribution to the late-time cosmic energy distribution, now referred to as  \textit{dark energy} \cite{spergel2007three}. The requirement for its existence has been established by various studies \cite{efstathiou2002evidence, spergel2003wmap}, but its nature remains one of the biggest mysteries in cosmology to date. Clarification of this problem is crucial, as dark energy is relevant to almost all fields of cosmology and is recognized as the overarching goal of precision cosmology \cite{wiltshire2008dark}. Dark energy not only dominates the expansion but also has substantial effects on the formation of highly nonlinear structure \cite{alimi2010imprints}, the cosmic microwave background (CMB) \cite{doran2001quintessence, caldwell2003early}, the dark matter distribution \cite{liberatorosenfeld2006darkenergy}, and the primordial structure growth \cite{ferreirajoyce1998scalingfield, bartelmann2005earlydarkenergy}.

Dark energy impacts the formation of large scale cosmic structures in ways that depend on its nature \cite{percival2005darkenergyandstructureformation, liberatorosenfeld2006darkenergy}. Thus, the cosmic history of structure formation is potentially an important probe of dark energy. This has been verified by \cite{alimi2010imprints} whom, using N-body simulations, have shown that uniformly distributed and constant-in-time dark energy leaves an imprint on galactic dynamics on all scales and most significantly in the nonlinear regime. 

It is curious that dark energy started to dominate the Universe's energy density around the time highly nonlinear structures on Megaparsec scales began to form, and especially that the dark energy density parameter is of the same order as the matter density parameter $\Omega_{m,0} \sim \Omega_{\Lambda,0}$ at present-time \cite{wetterich2002darkenergy, rasanen2004backreactionoflinear}, a phenomenon referred to as the \textit{coincidence problem}. This could simply be an extraordinary coincidence with no underlying physical explanation.  However, the coincidence begs the question whether the emergence of nonlinear structure could be the physical cause of the accelerated expansion. This has resulted in a plethora of investigations into the effect of nonlinear structures and on large-scale dynamics: the \textit{backreaction effect}, see e.g. \cite{buchert2000averagedust, brandenberger2000backreaction, wetterich2003can, rasanen2004darkenergy, kolb2005effectofinhomogeneities, coley2005macroscopicgravity, li2008darkenergyeffectofaveraging}. The earliest calculations of backreaction within the framework of general relativity were the calculations of the backreaction due to gravitational waves by Brill and Hartle \cite{brill1964method} and by Isaacson \cite{isaacson1968gravitational, isaacson1968gravitational2}. The question is whether there are any dynamical effects of cosmic structure on the expansion of the Universe, and whether these backreaction effects are sizable or not. Some claim that the backreaction is negligible \cite{ParanjapeSingh2008, green2011framework}, while others claim it could be causing the cosmic acceleration \cite{buchert2000averagedust, wiltshire2007cosmic}.

Note that nonlinear structure relates to the multiscaled nature of cosmic structure, and the central question can be regarded as determining dependencies of how the cosmic web evolves over different smoothing scales.

In this paper, we rigorously examine current prominent efforts in backreaction studies with respect to cosmic multiscaled nature. In Section \ref{section:current-methods} we show that the common perturbative approach to backreaction has implicit contradictory assumptions, making it impossible to consider a multiscale effects. We argue that an overlooked result on multiscaled effects refuting the common critique against the existence of backreaction, and that working with cosmic averages seemed to have been proven useful. To average over multiscaled structure we need a proper mathematical notion of how to slice our cosmology into spatial hyperspaces to average over, which has been a fundamental problem until now in many confusing and contradictory perspectives that have made backreaction studies problematic. To overcome this, we introduce a rigorous integral framework in Section \ref{section:foliation-gauges-in-rel-cosmology} on spacetime slicing and its relation to gauges in cosmology. In Section \ref{section:cosmic-averages-are-gauges} we show in what way standard spatial and 4-dimensional averages are implicitly gauge-dependent. Backreaction studies have used these gauge-dependent averages abundantly to quantify multiscaled properties of the large-scale structure, and we show by a simple example that this leads to artificial backreaction to the averaged acceleration of voids in Section \ref{sec:case-study}. To propel the study of multiscaled features of the cosmic web, we propose in Section \ref{sec:gauge-invariant-avg-proper-time-foliations} a solution to the gauge problem for averaging in relativistic cosmology for practical numeric usage. We show that the proposed method is gauge invariant. We conclude in Section \ref{sec:conclusion}.

In this paper, we solely consider classical gravity, without mention a general relativistic setting is thus assumed. If we do remark on the Newtonian case, we do so by explicitly stating this. Regarding notation, the Greek indices indicate the spacetime indices, whereas the Latin indices indicate the spatial ones.


\section{Current backreaction approaches subject to multiscaled cosmic structure}\label{section:current-methods}
An underestimated aspect of large-scale structure in backreaction studies is its multiscaled nature. We examine this characteristic in the context of two prevailing approaches: the Green-Wald perturbative approach \cite{green2011framework, green2012} and Buchert's averaging approach \cite{buchert2000averagedust, buchert2001averageperfectfluid}. The multiscaled structure reveals ambiguities within the theoretical framework of the perturbative approach and challenges the common ``Newtonian" critique against the averaging approach. This underscores the significance of the averaging approach, affirming its primary relevance in the current theoretical landscape for studying multiscaled large-scale structure.

\subsection{Perturbative approach subject to multiscaled structure}\label{subsec:current-methods:perturbative}

    Perturbative approaches are built upon realizing that inhomogeneities in the cosmic matter and energy distribution---such as clusters of galaxies---can be represented by deviations $\widetilde{g}_{\mu \nu}$, called \textit{perturbations}, of some smoothed-out metric $\overline{g}_{\mu \nu}$ which we employ as a background model:
    \begin{equation}\label{eq:perturbative-decomposition}
    g_{\mu \nu} = \overline{g}_{\mu \nu} + \widetilde{g}_{\mu \nu},
    \end{equation}
    see e.g. \cite{green2011framework, ishibashi2005acceleration}. Perturbative methods require a background model $\overline{g}_{\mu \nu}$ that contains the large-scale properties for appropriate backreaction results \cite{cliftonsussman2019backreaction}. Backreaction effects are, therefore, representable by the effects of the inhomogeneities $\widetilde{g}_{\mu \nu}$ on the background $\overline{g}_{\mu \nu}$, that is, inferring the dependence relation:
    \begin{equation}\label{eq:perturbative-backreaction}
    \overline{g}_{\mu \nu}(\widetilde{g}_{\mu \nu}).
    \end{equation}
    This is a critical drawback of the perturbative methods as identifying the large-scale properties in the real model $g_{\mu \nu}$ with an appropriate background $\overline{g}_{\mu \nu}$ that solves the Einstein field equations is not yet known \cite{clarkson2011growthstructure}. And even if suitable backgrounds are identified, the choice of background hugely affects the physical results \cite{ellis1987fitting}, in particular, the backreaction \cite{cliftonsussman2019backreaction}. This is immediate from \eqref{eq:perturbative-backreaction} as the choice $\overline{g}_{\mu \nu}$ determines the backreaction representation $\overline{g}_{\mu \nu}(\widetilde{g}_{\mu \nu})$.
    
    It is however of interest how dynamics in the perturbed model $\widetilde{g}_{\mu \nu}$ restrict the backreaction behaviour $\overline{g}_{\mu \nu}(\widetilde{g}_{\mu \nu})$. One of the present most prominent perturbative backreaction approaches is that of Green \& Wald \cite{green2011framework, green2012, green2013, green2014FLRW, green2016nobackreaction}. They pursue this by considering a smooth family $g_{\mu \nu}^{(\lambda)}$ of metrics parametrized by a scale $\lambda > 0$, where $g_{\mu \nu}^{(0)} = \overline{g}_{\mu \nu}$. With respect to several conditions, the backreaction then manifests itself as $t^{(0)}_{\mu \nu}$ in
    \begin{equation}\label{eq:perturbative-br-difference}
        G_{\mu \nu} ( \overline{g}_{\mu \nu} ) + \Lambda \overline{g}_{\mu \nu}=  \kappa \big( \overline{T}_{\mu \nu} + t_{\mu \nu}^{(0)} \big),
    \end{equation}
    where $t_{\mu \nu}^{(0)}$ is thought to be the energy-stress tensor of the backreaction \cite{green2016nobackreaction, bolejko2017current-status-inhomogeneous-cosmology}. Green \& Wald \cite{green2011framework} show that $t_{\mu \nu}^{(0)}$ is traceless and has positive energy density, $t_{\mu \nu}^{(0)} t^\mu t^\nu \geq 0$ for any timelike vector $t^\mu$ of the background. In particular, the backreaction does not contribute to dark energy.

    The cosmic web has different dynamical characteristics manifesting over its scales. For example, a small-scaled void seen locally on scales of 10 Mpc has a differently evolving velocity field than its overarching environment of densely packed voids eating one and another inside a, much larger, supervoid \cite{WeygaertKampen1993MNRASVoidsGravInstab}. It can therefore be argued that backreaction manifests over a range of scales \cite{buchert_ellis2015, ostrowski2017ongreenwaldformalism}, directly relating to the idea that macroscopic structure evolves differently than the sum of microscopic fluid particles. The multiscaled nature inherent in general relativity, and consequently the backreaction it entails, presents a challenge to one of the fundamental conditions of the perturbative approach as it should manifest over a whole range of scale parameter $\lambda$. An explicit derivation of this multiscaledness is however missing in the literature regarding the backreaction. We showcase this in light of the Green-Wald approach.
    
    Green \& Wald \cite{green2011framework} derive their formalism by assuming four conditions. Their Condition (i) states that the Einstein field equations hold for all scales $\lambda > 0$, that is,
    \begin{equation}\label{eq:green-condition-i}
        G_{\mu \nu} ( g^{(\lambda)} ) + \Lambda g_{\mu \nu}^{(\lambda)} =  \kappa T_{\mu \nu}^{(\lambda)},
    \end{equation}
    with $T_{\mu \nu}$ satisfying the weak energy condition with respect to $g_{\mu \nu}^{(\lambda)}$. This Condition is problematic in light of the multiscaled nature of the backreaction. 
    
    This issue is demonstrated through the non-commutative property of averaging and the derivation of the field equations, cf. Ellis et al. \cite{ellis1984relativistic, ellis1987fitting, ellis2005universe}. Suppose that at some fixed perturbation-scale $\lambda > 0$, the Einstein field equations \eqref{eq:green-condition-i} hold. We introduce a coarse-graining approach $\langle \cdot \rangle_\varepsilon$, which smooths out the cosmology represented by $g_{\mu \nu}^{(\lambda)}$ to an infinitely small order $\varepsilon > 0$. Here we restrict ourselves to a simplified cosmological model such that 
    \begin{equation}
        \big\langle G_{\mu \nu} (g^{(\lambda)}) \big\rangle_{2 \varepsilon} = \bigg\langle \big\langle G_{\mu \nu} (g^{(\lambda)}) \big\rangle_\varepsilon \bigg\rangle_\varepsilon,
    \end{equation}
    which can we understood as a restriction on the smoothness of structure on infinitesimal scales. Even in this restricted inhomogeneous model, the non-commutation of averaging implies the existence of a nonzero tensor\footnote{The best we can do is it to be a tensor field, which in that case would generally be nonzero. However it should be noted that it being a tensor field is not pre-determined. The mathematical fully nonlinear terms will most likely include structure functions.} field $t_{\mu \nu}^{(\lambda - \varepsilon)}$ such that:
    \begin{equation}
        G_{\mu \nu} \bigg( \big\langle g^{(\lambda)} \big\rangle_\varepsilon \bigg) = \bigg\langle G_{\mu \nu} \big( g^{(\lambda)} \big) \bigg\rangle_\varepsilon + t_{\mu \nu}^{(\lambda - \varepsilon)}.
    \end{equation}
    Given that the averaging was selected such that $\langle g_{\mu \nu}^{(\lambda)} \rangle_\varepsilon = g_{\mu \nu}^{(\lambda - \varepsilon)}$, we explicitly obtain a backreaction term at the scale $\lambda - \varepsilon < \lambda$:
    \begin{equation}\label{eq:pert-br-any-scale}
        G_{\mu \nu} \big( g^{(\lambda - \varepsilon)} \big) + \Lambda g_{\mu \nu}^{(\lambda - \varepsilon)} = \kappa \big( T_{\mu \nu}^{(\lambda - \varepsilon)} + t_{\mu \nu}^{(\lambda - \varepsilon)} \big).
    \end{equation}
    Here, we have utilized the relationship $\langle G_{\mu \nu}(g_{\mu \nu}^{(\lambda)}) \rangle_\varepsilon = \kappa T_{\mu \nu}^{(\lambda - \varepsilon)}$, assuming that the averaging operation is well-defined \cite{ostrowski2017ongreenwaldformalism, buchert_ellis2015}. This disproves the physical validation of Green-Wald's Condition (i) that assumes the field equations to be valid over all scales except at the background $\lambda=0$, where the backreaction is thought to be popping up. It shows that when smoothing out any degree of structure, an energy-stress backreaction term arises in the field equations.

    We indicate how the backreaction manifests over a whole multiscales in an infinitesimal scale $[\lambda - \delta, \lambda]$. When coarse-graining over multiple scales, say $n \in \mathbb{N}$ times with $n < \delta / \varepsilon$, the backreaction in \eqref{eq:pert-br-any-scale} can be seen to have a compounding effect,
    \begin{equation}\label{eq:avg-EFE-compounding}
        G_{\mu \nu} \big( g^{(\lambda - n\varepsilon)} \big) + \Lambda g_{\mu \nu}^{(\lambda - n\varepsilon)} = \kappa \big( T_{\mu \nu}^{(\lambda - n\varepsilon)} + t_{\mu \nu}^{(\lambda - n\varepsilon)} \big) + \kappa \sum_{i=1}^{n-1} \bigg\langle \cdots  \langle  t_{\mu \nu}^{(\lambda - i \varepsilon)}  \rangle_{i \varepsilon}    \cdots \bigg\rangle_{(n-1)\varepsilon},
    \end{equation}
    as $\langle t_{\mu \nu}^{(\lambda - i\varepsilon)} \rangle_{i\varepsilon}$ is generally nonzero. The repercussion of this compounding effect is the validity of Conditions (ii), (iii) and (iv) in Green \& Wald \cite{green2011framework}, which all relate to bounding the behaviour of $\widetilde{g}^{(\lambda)}_{\mu \nu}$ and its derivatives, which are governed by \eqref{eq:avg-EFE-compounding}. The multiscaled nature of the backreaction makes apparent that the inquiry should pertain to the limiting behavior of the averaging series in \eqref{eq:avg-EFE-compounding}, which is far from obvious, and generally does not satisfy such boundedness conditions. The limiting behaviour of the compounding effect in this averaging series could even be understood as the key question of backreaction studies.

    As analyzed above, the non-commutation of the Einstein field equations over the whole scale $\lambda\geq 0$ influences the validity of the Green-Wald conditions. Here we briefly show the effects on the conclusions of Green \& Wald as these conditions are not met in the cosmic web. In  particular, their result that the backreaction corresponds to solely gravitational waves with positive energy density, unable to affect dark energy. 
    
    The authors initialize their derivations by the relationship between the Ricci curvature $R_{\mu \nu}$ of $g_{\mu \nu}^{(0)}$ and $g_{\mu \nu}^{(\lambda)}$,
    \begin{equation}
        R_{\mu \nu} \big( g^{(0)} \big) - R_{\mu \nu} \big( g^{(\lambda)}  \big)  = 2 \nabla_{[\mu} {C^\rho}_{\rho]\nu} - 2 {C^\sigma}_{\nu[\mu} {C^\rho}_{\rho]\sigma},
    \end{equation}
    cf. Equations (9) and (10) of \cite{green2011framework}, with ${C^\rho}_{\mu \nu}$ defined with respect to $g_{\mu \nu}^{(\lambda)}$ and $\nabla_\mu$ the derivative operator on the background. The Einstein field equations on scale $\lambda$ give

    \begin{equation}\label{eq:green-riccicurvature}
        R_{\mu \nu} \big( g^{(0)} \big) - \Lambda g_{\mu \nu} = \kappa \big( T_{\mu \nu}^{(\lambda)} +  t_{\mu \nu}^{(\lambda)} \big) - \frac{\kappa}{2} g_{\mu \nu}^{(\lambda)} {T^\rho}_\rho^{(\lambda)} + 2 \nabla_{[\mu} {C^\rho}_{\rho]\nu} - 2 {C^\sigma}_{\nu[\mu} {C^\rho}_{\rho]\sigma},
    \end{equation}
    cf. Equation (19) in \cite{green2011framework}, including the backreaction term $t^{(\lambda)}_{\mu \nu}$ at scale $\lambda$ however. The authors then take the weak-limit $\lambda \to 0$ of $\nabla_{[\mu} {C^\rho}_{\rho]\nu}$ to retrieve $t_{\mu \nu}^{(0)}$. Excluding the backreaction term, the weak-limit of the Ricci curvature expression \eqref{eq:green-riccicurvature} then gives rise to traceless $t_{\mu \nu}^{(0)}$. The integral problem of the multiscaled nature of the backreaction manifests itself that the weak-limit of $t_{\mu \nu}^{(\lambda)}$ is ambiguous, and might not even exist \cite{buchert2012backreaction}. The backreaction on the background $t_{\mu \nu}^{(0)}$ will in general therefore not be traceless, and hence encapsulate effects different than gravitational waves.
    
    Even if there exists a weak-limit $\text{w}\lim_{\lambda \to 0} t_{\mu \nu}^{(\lambda)}$ of the backreaction term, then this would in general be nonzero. And, thus 
    \begin{equation}\label{eq:green-weaklimit}
        \text{w}\lim_{\lambda \to 0} \; \widetilde{g}_{\rho \sigma}^{(\lambda)} \nabla_{[\mu} {C^\rho}_{\rho]\nu} = - \frac{\kappa}{2} \; \text{w}\lim_{\lambda \to 0} \; \widetilde{g}_{\rho \sigma}^{(\lambda)} \bigg( T_{\mu \nu}^{(\lambda)} +  t_{\mu \nu}^{(\lambda)}  - \frac{1}{2} g_{\mu \nu}^{(\lambda)} {T^\rho}_\rho^{(\lambda)} \bigg),
    \end{equation}
    is nonzero, cf. Equations (20) up to  (30) of \cite{green2011framework}. Thus, generally, $t_{\mu \nu}^{(0)}$ has a nonvanishing trace. This also troubles the derivation of $t_{\mu \nu}^{(0)}$ being a positive energy density is a direct consequence of the backreaction term being traceless, cf. Equation (31) up to (34) of \cite{green2011framework}. 

    We see that when including the existence of multiscaled structure, and thus its geometrical effects, contradicts all most all of the fundamental assumptions made by the Green-Wald perturbative approach to backreaction. And it follows from the above consecutive arguments that their results on the non-existence of large-scale backreaction are thereby refuted. Other reevaluations of the Green-Wald approach are found in for example \cite{buchert_ellis2015, clifton2019backreactionSphericalSpacetimes}.


\subsection{Averaging approach subject to multiscaled structure}\label{subsec:current-methods:averaging}
We argued that the multiscaled nature of cosmic structure, and thus the backreaction, forms inherent problems to the current perturbative formalism. On the other hand, the averaging approach to backreaction has gotten significant attraction.

The averaging approach covers and divides an inhomogeneous cosmological model into compact regions and quantify local dynamics in each region by averaging over matter fields, such as the energy density, describing the content of the Universe; see e.g. Zalaletdinov \cite{zalaletdinov1992averaging, zalaletdinov1993towards, zalaletdinov1997averaging, zalaletdinov2008averaging} and Buchert \cite{buchert2000averagedust, buchert2001averageperfectfluid}. For reviews, see \cite{buchert2008dark, wiltshire2008dark, wiltshire2009gravenergy, wiltshire2011dust, ellis2009dark, buchert2012backreaction}. To the best of our knowledge, averages in general relativity and cosmology have been introduced by Shirokov \& Fisher \cite{Shirokov1963NonhomogeneousUniverse} and Ellis \cite{ellis1984relativistic}; also see \cite{noonan1984gravitational, zotov1992averaging, zotov1995averaging} and \cite{futamase1988, futamase1989}. Note that averaging over the perturbed metric means smoothing the (local) cosmic inhomogeneities, which represent the complex dynamics and structuring of cosmic matter.

Quantitative contributions using the averaging scheme present results that seem to indicate the backreaction to be significantly influencing cosmological parameters \cite{kolb2005effectofinhomogeneities, kolb2006acceleration, enqvist2007ltb, marozziuzan2012anisotropy, adamek2015affectdynamics}. Examples are Buchert et al. \cite{buchert2000cosmparamaters} portraying that the cosmological density parameters, such as $\Omega_\Lambda$, in a region of 100 Mpc possibly can deviate more than 100 percent for $3\sigma$-fluctuations in the initial cold dark matter density; and Räsänen \cite{rasanen2004darkenergy} portraying that the backreaction can result in an equation of state $0 \leq \omega \leq -4/3$ without the need for dark energy. Even if backreaction effects are small for sufficiently large regions, perturbations can result in large deviations from the standard homogeneous values of the cosmological parameters on scales of 100 Mpc or less \cite{buchert2000cosmparamaters}. Since even small deviations in the cosmic matter distribution can lead to significant changes to the macroscopic description of a physical system \cite{ellis2005universe}, the above results imply a realistic possibility of the backreaction effect explaining the cosmic acceleration.

The above results on the backreaction effect have been criticized by those presenting perturbative arguments \cite{green2011framework, green2014FLRW, green2016nobackreaction} and Newtonian arguments. The former has been commented on above. The latter emerges from the idea that the local Universe is accurately described by Newtonian mechanics, since the gravitational potential is small, all velocities are small and all scales are small compared to the Hubble length $cH$ \cite{peebles1980large}. The backreaction appears to be nonexistent \cite{kaiser2017there, buchert2018newtonianbackreaction}, or at least negligible \cite{ishibashi2005acceleration, green2014FLRW}, in Newtonian theory. If backreaction effects are sizable, they must appear on Newtonian scales in the nonlinear regime, and thus especially in the local Universe \cite{ishibashi2005acceleration}. This has led to the conclusion that there is no backreaction on global scales---not even in the relativistic setting, cf. Kaiser \cite{kaiser2017there}. Up to date, this appears to have confirmed the widely shared view on the backreaction within the scientific community: its effects are negligible on large-scale cosmic dynamics.

In recent years, it has become apparent that Newtonian dynamics does not adequately describe the local Universe in all relativistic scenarios, despite the gravitational potential, velocities and scales involved being small and previously thought to be Newtonian. For example, Korzyński \cite{korzynski2015nonlineareffectsofgr} rigorously shows that there are nonlinear effects caused by the way cosmic structure is nested and hierarchically distributed, and that these cannot be included within Newtonian gravity or first-order relativistic perturbation theory. 

Furthermore, it is shown that these effects can be very significant as the accumulation of nonlinear effects in general relativity can amplify backreaction of inhomogeneities---if only enough different orders of scales are considered. The backreaction can consequently have sizable impacts on physical situations we have relied on to be almost perfectly Newtonian, ranging from structure formation and $N$-body simulations to determining the Hubble constant \cite{racz2017withoutDarkEnergy, wiltshire2008dark}.

The most prominent formalism up to now for describing backreaction effects on large-scale cosmic dynamics, and in particular the multiscale effects due to the clustering and nesting of structure, seems to be that of averaging. In the following sections, we provide the mathematical background of cosmic averaging with respect to the spacetime slicing, from which we identify a fundamental problem.

The question of constructing an appropriate averaging method arises naturally when dealing with cosmic dynamics. However, in the case of a curved space, averaging is not geometrically defined as there is not a unique way to average nonlinear fields, cf. \cite{buchert_Ehlers1995averaging, gasperini2009gauge, brannlund2010averaging}. Below, we show that averaging turns out to implicitly induce gauges. This presents a challenge in accurately averaging the multiscaled dynamics of the cosmic web.


\section{Foliation gauges in relativistic cosmology}\label{section:foliation-gauges-in-rel-cosmology}
As discussed above, the literature suggests that averaging holds promise as a method for describing the cosmic multiscaled effects---possibly on the expansion. The mathematical procedure of averaging in cosmology is fundamentally linked to the notion of slicing the spacetime manifold, whose complex nature has caused much confusion in the literature. Specifically, we are interested in claims made regarding gauges in backreaction studies; different notions of and claims about gauge invariance in the literature, compare for example \cite{ishibashi2005acceleration, buchert2012backreaction, mourier2024splittingspacetime}. Our treatment below elucidates this formally. In particular, we provide a mathematical treatment of a slicing of spacetime, also called a \textit{foliation}, for cosmology. We lay out the correspondence between foliations and cosmological gauges by presenting an integral framework for foliation-gauge transformations in relativistic late-time cosmology.

\subsection{General relativity as gauge theory}
    We state this section for completeness, the rest the paper should also be accessible to those who do not have a background in mathematical gauge theory. For for an excellent introduction to fibre bundles for physicists see Marsh \cite{marsh2016gaugetheories}.

    In the mathematical literature it is well-known that GR is a gauge theory, which roughly means that it is a physical theory described on a $G$-principal bundle $P \xrightarrow{\pi} M$ with structure group $G$ and $\pi$ its projection onto $M$, where the system's dynamics are governed by a principal connection \cite{trautman1980fiber}. An example is the frame bundle $FM \xrightarrow{\pi} M$ over general linear group $\gl$, where $FM$ is the unique space of all frames on $M$, \textit{i.e.} ordered bases $\{f_\mu\}$ with $f_\mu \in TM$. There exists a unique Levi-Civita connection $\nabla$ on the tangent bundle of $M$, which can be identified with a unique principal connection on its frame bundle \cite{michor2008topics}. This principal connection, in turn, induces the Levi-Citiva connection. We can thus fully describe GR on the frame bundle of $M$.
    
    Any fibre $F_p M$ is diffeomorphic to $\gl$ by $L \mapsto f \cdot L$ for any fixed $f_p \in F_p M$. By definition of a principal bundle, locally there exists some $U\subset M$ and a diffeomorphism $\phi: FM |_U \xrightarrow{~} U \times \gl$ such that for any frame $f \in FM|_U$,
    \begin{equation}
        \phi(f \cdot L) = \phi(f) \cdot L \qquad \text{for all} \; L\in \gl,
    \end{equation}
    where in the right-side $\gl$ only acts on the second component. Since $FM|_U \cong U \times \gl$ via $\phi$, we call it a local trivialization or a \textit{(local) gauge}. There is an one-to-one correspondence between local gauges $\phi: FM|_U \xrightarrow{~} U \times G$ and smooth maps $f:U \to FM|_U$, called local sections, which are therefore also referred to as local gauges. Note that a coordinate system $x$ induces the local gauge $\{ \partial_{x^\mu} \}$.
    
    The spacetime symmetry\footnote{GR does not yield `internal symmetries' as there is no $\Gamma \in \diff(FM)$ that transforms the `internal space' $F_x M$ and leaves the underlying spacetime $M$ unchanged. Physicists sometimes require a gauge theory to have an internal symmetry, explaining why some claim GR is not a gauge theory.} is the group of \textit{gauge transformations} $\mathcal{G}$ are the diffeomorphisms $\diff(FM)$ that preserve the solder form on $FM$. It can be proven that $\mathcal{G} \cong \diff(M)$ cf. \cite{trautman1980fiber}. The linearization at the identity, so to say the infinitesimal approximation, of $\diff(M)$ is described by its Lie algebra, which is generated by infinitesimal diffeomorphisms $x^\mu \mapsto x^\mu + \xi^\mu$. Any manifold's description is independent of local coordinates, that is the equations should be invariant under coordinate transformation $x \mapsto \widetilde{x} = L(x)$ for any $L \in \gl$, the so-called general covariance.

\subsection{Relativistic cosmology as gauge theory}\label{subsec:cosmology-as-gauge-theory}
    The mathematical model of cosmology is given by a Lorentzian manifold with tensor fields $T: M \to \mathcal{T}M$, called matter fields, representing classical particles and its characteristics \cite[Ch.~3.2]{hawkingellis1973}. Any gauge transformation $\gamma \in \diff(M)$ affects the full model description $\big(M,g,\{F\} \big)$, abbreviated by $M$, in the following way: 
    \begin{equation}\label{eq:GT}
        M \ni p \longmapsto \gamma(p), \qquad g \longmapsto \gamma_* g, \qquad F \longmapsto \gamma_* T.
    \end{equation}
    For infinitesimal $\gamma \in \diff^*(M)$ we have $\gamma_* T = T + \mathcal{L}_\xi T$ with $\xi$ the infinitesimal generator of $\gamma$. Since $\diff(M)$ is the spacetime symmetry of $M$, it thus generates the equivalence class of all cosmological models $\big[ (M, g, \{T\} ) \big]$ under \eqref{eq:GT} that are physically equivalent as they describe the same physical reality, cf. Hawking \& Ellis \cite[Ch.~3.1]{hawkingellis1973}.\footnote{Note that \cite[Ch.~3.1]{hawkingellis1973} call any $\gamma \in \diff(M)$ an `isometry' as its leaves the line element $g_{ab}(x)dx^a dx^b$ is invariant. Usually, however, we reserve the term for the subgroup $\text{Iso}(M) \subset \diff(M)$ as described in Subsection \ref{subsec:gaugetransformationsoffoliations}.} Since
    \begin{equation}
        \big[ (M, g, \{T\}) \big] = \big\{ (M, \gamma_* g, \gamma_* \{T\}) \mid \gamma \in \mathcal{G} \big\} \cong \mathcal G,
    \end{equation}
    a representation $(M, g, \{T\})$ from its equivalence class can be called a \textit{gauge}.

    \begin{figure}[h!]
        \centering
        \includegraphics[width=17cm]{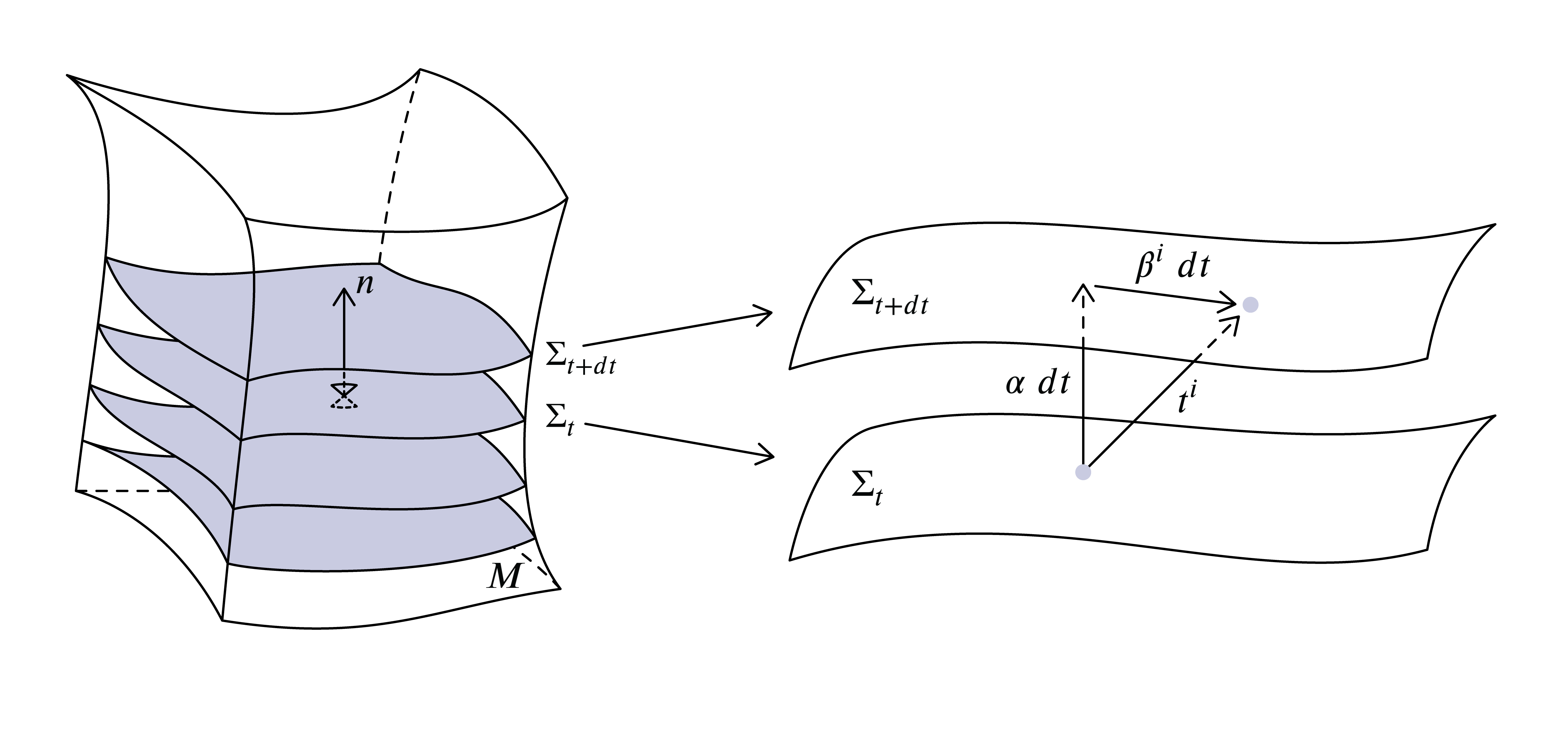}
        \caption{A depiction of a foliation with spatial hypersurfaces $\Sigma_t := F_t(\Sigma)$ in terms of its shift and lapse variables.}
        \label{fig:foliationvariables}
    \end{figure}

    The most common notion of gauges in cosmology are those found in perturbation theory. To see this in context of the above, pick a gauge $\b{M}$ from the equivalence class of $M$ with $\b{\gamma}: \b{M} \to M$ the unique gauge transformation between the two models. Here $\b{M}$ specifies the background model, but of course $\gamma(\b M) \cong M$. Fixing a corresponding background field $T^{(0)}: \b M \to \mathcal{T}M$, we can globally decompose its matter field $T: M \to \mathcal{T}M$ as
    \begin{equation}
        T(p) = T^{(0)} \big( \b \gamma^{-1}(p) \big) + \delta T (p),
    \end{equation}
    where perturbation $\delta T$ is generally dependent on $\gamma$ --- or as cosmologists say, the choose of background $\b M$, and thus gauge-dependent.
    
    Gauge transformation $\gamma: \b{M} \to M$ thus generates a perturbation transformation. Take $\gamma$ to correspond to infinitesimal generator $\xi^\mu$, then the transformed perturbation is $\widetilde{\delta T} = \delta T + \mathcal{L}_\xi T$. Locally $\gamma$ can be represented as the infinitesimal coordinate transformation $x^\mu \mapsto x^\mu + \xi^\mu$. This framework thus captures the active and passive approach to gauge transformations in perturbation theory, cf. \cite[Ch.~3]{mukhanov1992theory}. Note that a random pick of $\widetilde{\delta T}$ such that  $\delta T \mapsto \widetilde{\delta T}$ need not be corresponding to some $\gamma \in \diff(M)$, and in that case formally cannot be understood as a gauge transformation.\footnote{For completeness we recall that spacetime $M$ is a Lorentzian $4$-manifold with a \textit{time-orientation} $\mathfrak{t}:M\to TM$, \textit{i.e.} $g_p(\mathfrak{t}_p, \mathfrak{t}_p) < 0$ for all $p \in M$. All gauge transformations of spacetime are required to preserve the time-orientation, and thus the manifold-orientation. We thus only consider such transformations throughout the rest of the paper.}
    
    Relating to literature on the backreaction, we contrast this to the discussion in e.g. Buchert \& Räsänen \cite[Sec.~2.2]{buchert2012backreaction}, where perturbation-gauge transformations $\delta T \mapsto \widetilde{\delta T}$, coordinate transformations $x^\mu \mapsto \widetilde{x}^\mu$ and foliation transformation $F \mapsto \widetilde F$ (discussed below) are seen as independent constructions. The authors state that their averaging procedure is `gauge-invariant', and solely dependent on a choice of foliation over which they average. The above rigorous formalism for cosmological gauges, however, shows the intricate nature of these related concepts. The rest of the paper is devoted to make the notion of cosmic foliations and its relation to gauges explicit.

\begin{figure}[h!]
    \includegraphics[width=0.4\textwidth]{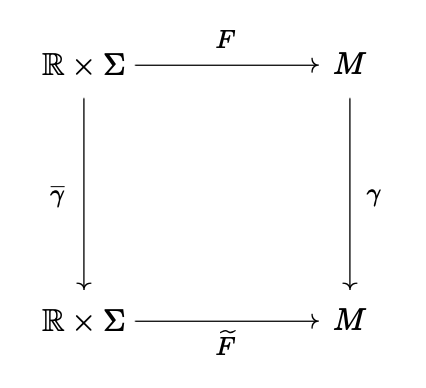}
    \centering 
    \caption{Commutative diagram of foliation-GT denoted by $F \mapsto \widetilde{F}$.}
    \label{fig:commdiagram}
\end{figure}

\subsection{Gauge transformations of spacetime foliations}\label{subsec:gaugetransformationsoffoliations}
It is known that splitting spacetime $M$ is a matter of choice, although a formal perspective on this is missing for cosmological backreaction. The outlined holistic gauge-theoretical approach however gives an appropriate context. We discuss several relevant examples in the subsequent subsection.

A \textit{foliation} of $M$ is a diffeomorphism $F:\R \times \Sigma \xrightarrow{~} M$ such that $\Sigma_t := F_t(\Sigma)$ is a spacelike hypersurface. There exists a foliation, that is $M$ is globally hyperbolic, if and only if there exists a scalar $\t: M \to \R$, called a temporal function, such that $g(\nabla \t, \nabla \t) < 0$ everywhere on $M$. If so, we can define the lapse function $\alpha$ and shift vector $\beta^i$ in a coordinate-independent way by
\begin{equation}\label{eq:foliation-GT}
    \alpha := \frac{1}{\sqrt{-g(\nabla \t, \nabla \t)}},
    \qquad
    \n := - \alpha \nabla \t,
    \qquad
    \boldsymbol{\beta} := \frac{dF_t}{dt} - \alpha \n,
\end{equation}
where $\n$ is the unit normal to $\Sigma_t$, see e.g. \cite{landsman2021foundations} and Figure \ref{fig:foliationvariables}. Since $F_t:\Sigma \to M$ is an embedding, it induces the spatial metric $h:= F_t^* g$ on $\Sigma_t$. Note that a random choice of $\alpha$ and $\beta^i$ does not generally define a foliation $F$.

\begin{figure*}[h]
    \centering
    \begin{subfigure}[t]{0.5\textwidth}
        \centering
        \includegraphics[width=8.5cm]{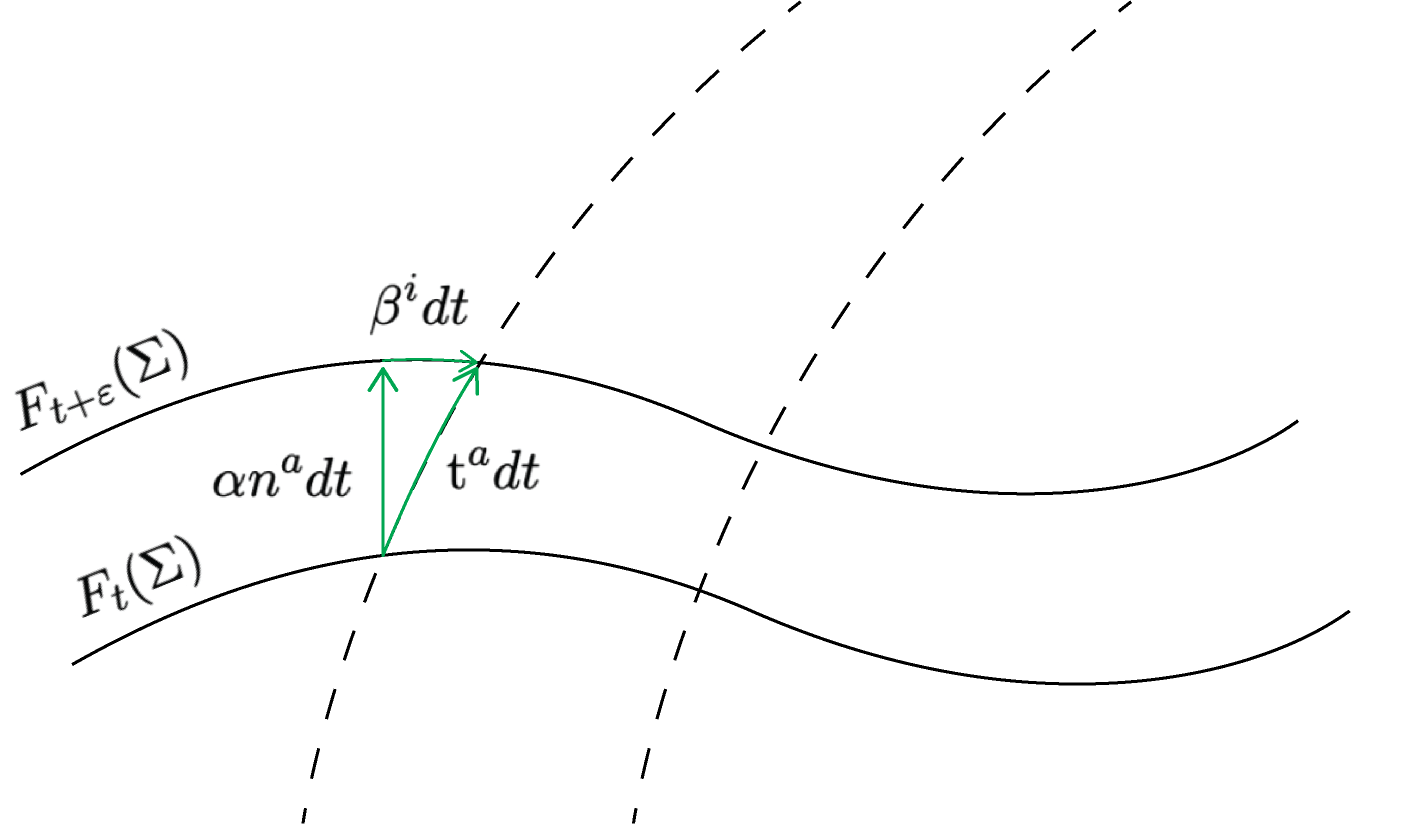}
        \caption{Foliation $F$ in green.}
    \end{subfigure}%
    ~ 
    \begin{subfigure}[t]{0.5\textwidth}
        \centering
        \includegraphics[width=8.5cm]{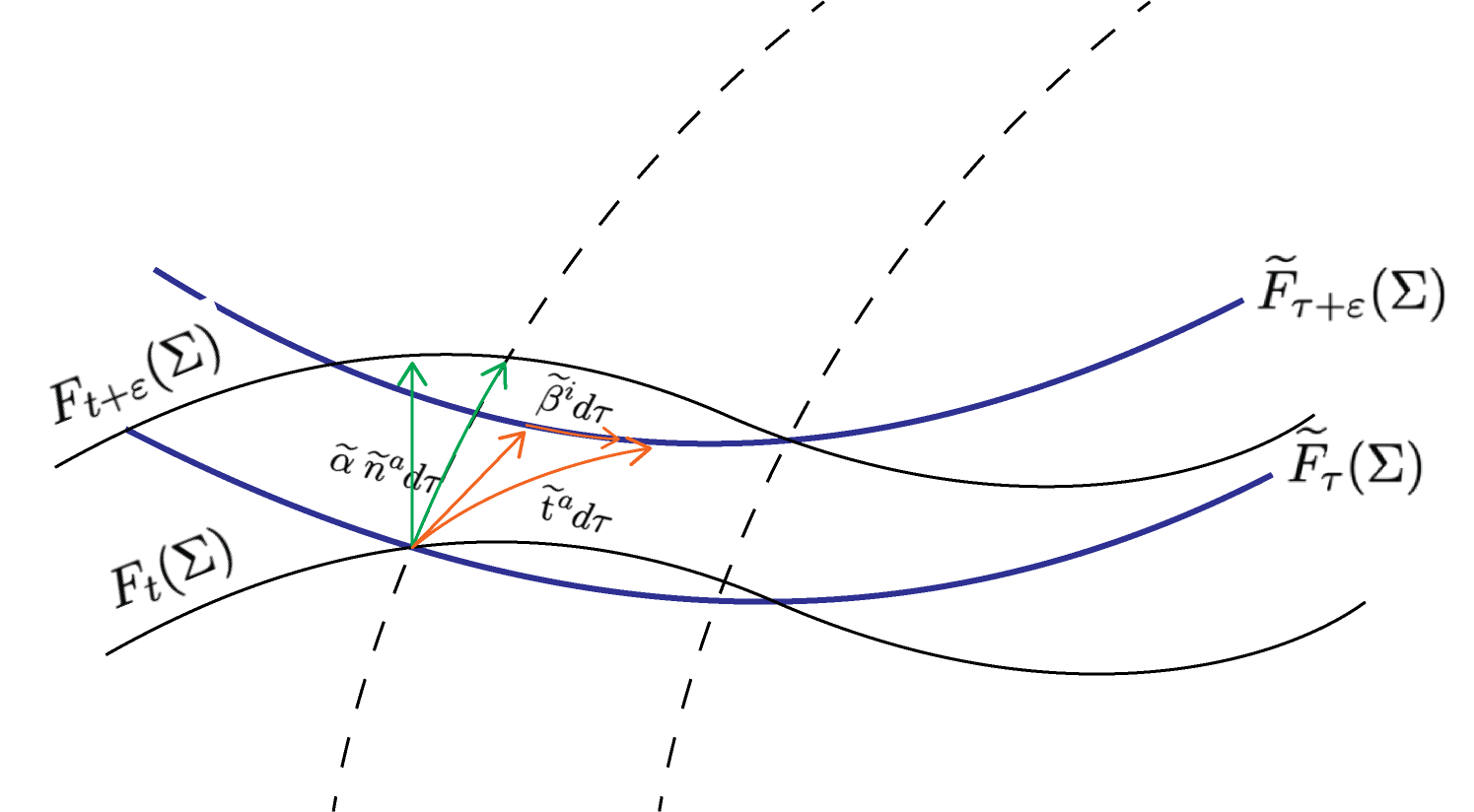}
        \caption{Transformation $F \to \widetilde{F}$ in orange.}
    \end{subfigure}
    \caption{Sketches of foliation change $F \mapsto \widetilde{F}$ in terms of the foliation variables.}
    \label{fig:foliationGT}
\end{figure*}

Foliations are gauges as foliation changes are gauge transformations. To see this correspondence, a foliation transformation $F \mapsto \widetilde{F}$ induces the unique gauge transformation $\gamma:= \widetilde{F} \circ F^{-1} \in \diff(M)$, which is orientation-preserving as the casual structure is preserved. One might also consider to keep $(M,g)$ fixed, so set $\gamma=\text{id}$, then this would yield $\b{\gamma} = \widetilde{F}^{-1} \circ F \in \diff(\R \times \Sigma)$. This corresponds to the fact that one is able to formulate different foliations on a specific representation $(M,g) \in [(M,g)]$. Reversely, given a foliation $F$, any $\gamma \in \diff(M)$ induces the foliation transformation $F \mapsto \gamma \circ F$ as $\gamma^* F_t (\Sigma)$ is spacelike since $\gamma$ is orientation-preserving. Similarly, $F \mapsto \b \gamma^* F$ for any $\b \gamma \in \diff(\R \times \Sigma)$. For all these scenarios, the commutative diagram in Figure \ref{fig:commdiagram} commutes. Under a gauge transformation $\gamma \in \diff(M)$, the foliation transforms as
\begin{equation}\label{eq:GT-foliation-functions}
    \alpha \longmapsto \gamma_* \alpha,
    \qquad 
    \boldsymbol{\beta} \longmapsto \gamma_* \boldsymbol{\beta},
    \qquad
    h \longmapsto \gamma_* h,
\end{equation}
which we depict in Figure \ref{fig:foliationGT}.

\subsection{Several examples}
Here we discuss several examples of the foliation-gauge transformations outlined above. They naturally relate to known kinds of diffeomorphisms due to the identification relation as became clear from understanding cosmology in light of mathematical gauge theory.
\begin{itemize}
   \item[(a)] \textit{Time-reparametrizations}. 
                A simple example of a foliation-gauge transformation is a global reparametrization of the time-parameter $t \mapsto \tau \in \R$ of $F_t$ given by $\gamma(t,\boldsymbol x) = \big( \tau (t), \boldsymbol{x} \big)$. Then the time-component of $\nabla \t$ transforms as
                \begin{equation}
                    g^{00} \frac{\partial \t}{\partial x^0} \longmapsto dt{\tau}^2 g^{00} \frac{\partial \t}{\partial x^0},
                \end{equation}
                and thus according to \eqref{eq:foliation-GT} we have $\gamma_* \alpha = \alpha / dt{\tau}^2$, and $n^\mu, \beta^i, h_{ij}$ are invariant. Cosmic averaging invariant under time-reparametrizations are treated in a rigorous manner by a recently proposed framework of Mourier \& Heinesen \cite{mourier2024splittingspacetime}. With the formalization of foliation-gauges in cosmology we however see that time-reparametrizations are just a specific form of foliation-gauge transformations.

    \item[(b)] \textit{Conformal transformations}.
                Another example of gauge transformations are global conformal transformations, which are diffeomorphisms $\gamma$ such that $\gamma^* g = \Omega^2 g$ for some positive scalar $\Omega: M \to \R$, \textit{i.e.} they preserve the geometry's angles. Locally, we have $x^\mu \mapsto \Omega(x) x^\mu$, and thus $g_{\mu \nu} \mapsto \Omega^2 g_{\mu \nu}$. It transforms the foliation as
                \begin{equation}\label{eq:GT-foliation-conformal}
                    \alpha(x) \longmapsto \Omega(\tilde x) \alpha(\tilde x), 
                    \qquad
                    \beta_i(x) \longmapsto \Omega(\tilde x) \beta_i(\tilde x)
                    \qquad
                    h_{ij}(x) \longmapsto \Omega^2(\tilde x) h_{ij}(\tilde x).
                \end{equation}
                It thus leaves no slicing parameters invariant in contrast to time-reparametrizations. Note that the conformal transformations form a subgroup $\text{Conf}(M) \subset \diff(M)$. 

    \item[(c)] \textit{Isometries}.
                Global scale transformations are conformal transformations that assign a fixed scalar value $\Omega(x) = \Omega$ to every point $x\in M$, \textit{i.e.} they preserve lengths. A special case of these are transformations $\gamma \in \diff(M)$ such that $\Omega=1$, called isometries; for which we have
                \begin{equation}
                    g_{\mu \nu}(x) \longmapsto \widetilde{g}_{\mu \nu}(\tilde x) = \frac{\partial x^\rho}{\partial \tilde x^\mu} \frac{\partial x^\sigma}{\partial \tilde x^\nu} g_{\rho \sigma}(x) = g_{\mu \nu}(\tilde x),
                \end{equation}
                forming the subgroup $\text{Iso}(M) \subset \diff (M)$. The foliation transforms according to \eqref{eq:GT-foliation-conformal},  under which the slicing is not invariant for non-trivial isometries.
\end{itemize}

\subsection{Relation between foliations and cosmological perturbation gauges}\label{subsec:relation-between-foliations-and-perturbation-theory}
The full formulation of gauges in cosmology in terms of mathematical gauge theory made the theoretical identification between the gauges in cosmological perturbation theory and spacetime diffeomorphisms. Decomposing foliation $F$ shows this identification. The general decomposition of in terms of foliation-gauge $F_t$ is
\begin{equation}
    g_{\mu \nu} dx^\mu dx^\nu = (\beta_i \beta^i - \alpha^2)dt^2 + 2\beta_i dx^i dt + h_{ij} dx^i dx^j.
\end{equation}
In conformal time $\eta$, the full-order perturbed FLRW metric is given by
\begin{equation}
    g_{\mu \nu} dx^\mu dx^\nu = a^2(\eta) \big( -(1+\phi) d\eta^2 + 2B_idx^i d\eta + (\delta_{ij} + 2C_{ij}) dx^i dx^j\big), 
\end{equation}
see e.g. \cite{Brown2009JCAP:GaugesAndBackreaction}. The foliation-gauge in conformal time $t=\eta$ can thus be expressed in terms of the standard perturbation gauge variables by
\begin{equation}
    \beta_i = B_i, \qquad
    h_{ij} = \delta_{ij} + 2 C_{ij}, \qquad
    \alpha^2 = 1 + \phi + (\delta_{ij}+2C_{ij})B^i B^j.
\end{equation}

The common gauges found in cosmological perturbation theory are easily captured and translated to the conditions unto the corresponding foliation. For example, the `uniform curvature gauge' are foliations of the form
\begin{equation}
    \beta_i = B_i, \qquad
    h_{ij} = \delta_{ij}, \qquad
    \alpha^2 = 1 + \phi + B^i B^i.
\end{equation}
And the `conformal Newtonian gauge' are foliations of the form
\begin{equation}
    \beta_i = 0, \qquad
    h_{ij} = (1 - 2 \Phi)\delta_{ij}, \qquad
    \alpha^2 = 1 + \phi.
\end{equation}


\section{Cosmic scalar averages are gauges as they depend on foliations}\label{section:cosmic-averages-are-gauges}
With the formal treatment of foliations laid down above, we elucidate its role in cosmic averaging. This treatment brings to light a fundamental problem of these averaging procedures: selecting an averaging method necessitates specifying a spacetime-slicing, for which averaging is not invariant under its transformations. Consequently, the choice of averaging is not inherent to the cosmological model; that is, they are gauges. This limitation renders standard averaging non-representative for structure analyses in relativistic cosmology.

\subsection{Foliation gauge transformations over 4-dimensional volumes}
    Consider a 3-dimensional compact domain $\D \subset \Sigma$ and pick a foliation $F(t,x)$. We are interested in the subspace $F(T \times \D) := \cup_{t \in T} F_t(\D) \subset M$ generated by foliation $F_t$ evolving $\D$ over the time-interval $T \subset \R$. Usually, one implicitly defines $F_t(\D)=\D_t \subset \Sigma_t$. Since $F(T \times \D)$ is 4-dimensional, the integration over scalar $S:M \to \R$ is formulated through integrating over the volume\footnote{We write $dx^3 = dx^1 \wedge dx^2 \wedge dx^3$ and $dtd^3x = dt \wedge d^3x$.} 4-form $\S := S(t,y) \sqrt{-g(t,y)} \; dt d^3y$,
    \begin{equation}
        \int_{F(T \times \D)} \S 
        = \int_{T \times \D} F^* \S 
        = \int_{T \times \varphi_x(\D)} S \big( F(t,x) \big)  \sqrt{-g\big( F(t,x) \big)} \; dt d^3x
    \end{equation}
   emphasising that $\varphi_x$ is a coordinate\footnote{The coordinate map $\varphi_x:U \to \mathbb{R}^3$ is defined on chart $U \subset \Sigma$. We implicitly understand this to represent any set of coordinates with charts covering and adapted to $F_t(D)$ as can be made explicit with a partition of unity. The coordinates must however be adapted to foliation $F$.} map on $\Sigma$ with respect to foliation $F_t$. We can however also pick coordinates $(t,y)$ on $M$ such that we retrieve the standard integral expression
    \begin{equation}
        \int_{F(T \times \D)} \S 
        = \int_{\varphi_{t,y} (F(T \times \D))} S(t,y)  \sqrt{-g(t,y)} \; dt d^3y,
    \end{equation}
    which is the standard form used in computation especially in backreaction studies. It thus constitutes an implicit foliation-gauge choice.
    
    Under a foliation-gauge transformation $F \mapsto \gamma \circ F$, the corresponding integral transforms as
    \begin{equation}\label{eq:GT-4integral}
        \int_{F(T \times \D)} \S 
        \longmapsto 
        \int_{\gamma \circ F(T \times \D)} \S  
        = \int_{ F(T \times \D)} \gamma^* \S
        = \int_{ \varphi_{t,y}(F(T \times \D))} S \big( \gamma (t,y)\big) \sqrt{-g \big(\gamma(t,y) \big)} \; dt d^3 y,
    \end{equation}
    with $\varphi_{t,y}$ coordinates on $M$, and where the first two integral expressions highlight the transformation of the integral domain. In components, the transformed integral is
    \begin{equation}
        \int_{F(T \times \D)} \S 
        \longmapsto 
        \int_{ T \times \D}F^*  \gamma^* \S
        =
        \int_{T \times \varphi_x(\D)} S \big(\gamma \circ F(t,x) \big) \sqrt{-g \big(\gamma \circ F(t,x)\big)} \; dt d^3x,
    \end{equation}
    with $\varphi_x$ coordinates on $\D \subset \Sigma$. It points out the effect of the gauge-transformation on the spacetime structure. As viewed from these different perspectives, the integral and the corresponding 4-dimensional average is, thus, generally not gauge-invariant, cf. Gasperini et al. \cite{gasperini2009gauge}.

\subsection{The averaging dependence on foliation choice}\label{subsec:averaging-dependence-on-foliation-choice}
    It is standard practice to work with a corresponding spatial integral over each 3-dimensional time-slice $F_t(D)$, where the correspondence 3-form is taken to be $S(x) \sqrt{h(x)} d^3 x$. This form is however rather suggestive as it implicitly depends on the foliation. To be precise, we naturally induce a 3-form on the spatial hypersurface $\Sigma_t$ by the interior product $\iota_{\n} \S$ along the timelike vector $\n=\alpha \nabla \t$. The integral over the induced 3-form $\S$ on the spatial hypersurface $F_t(\D)$ is then
    \begin{equation}\label{eq:spatial-avg-comp}
        \int_{F_t (\D)} \iota_{\n} \S 
        = \int_{\varphi_y ( F_t (\D) )} S(t,y) \sqrt{-g(t,y)}  dt(\n) \; d^3 y
        = \int_{\varphi_y ( F_t (\D) )} S(t,y) \sqrt{h(t,y)}  \; d^3 y,
    \end{equation}
    as $\sqrt{-g(t,y)} = \alpha \sqrt{h(t,y)}$ and $dt(\alpha \n) = 1$. Here we retrieve the standard coordinate expression of the spatial integral by considering $\varphi_{t,y}$ coordinates on $\Sigma_t \subset M$. We can however also express it with the foliation term in the integrand as follows:
     \begin{equation}
        \int_{F_t (\D)} \iota_{\n} \S 
        = \int_{\D} F_t^* \big( \iota_{\n} \S \big)
        = \int_{\varphi_x(\D)} S \big( F(t,x) \big) \sqrt{h \big( F(t,x) \big)} \; d^3x,
    \end{equation}
    where $\varphi_x$ are now coordinates for $\D \subset \Sigma$. Elucidating the foliation dependence as found \eqref{eq:spatial-avg-comp}.
    
    A gauge-transformation works on $M$ according to \eqref{eq:GT-4integral}, after which one induces the 3-form on the foliation slice.\footnote{Note that gauge transformations and spatially inducing volume forms on the foliation slices do not commute, since first inducing and then transforming the foliation gives $\int_{\gamma \circ F_t(\D)} \iota_{\n} \S = \int_{F_t(\D)} \iota_{\gamma^* \n} \gamma^* \S$.} Under $F \mapsto \gamma \circ F$ with its lapse and shift transforming as \eqref{eq:GT-foliation-functions}, the spatial integral transforms according to
    \begin{equation}\label{eq:spatially-induced-avg}
        \int_{F_t (\D)} \iota_{\n} \S 
        \longmapsto 
        \int_{\gamma \circ F_t(\D)} \iota_{\gamma_* \n} \S
        .
    \end{equation}
    Since $dt(\alpha \n)$=1 and $\sqrt{-g} = \alpha \sqrt{h}$, that is in components,

    \begin{figure*}[t!]
    \centering
    \begin{subfigure}[t]{0.5\textwidth}
        \centering
        \includegraphics[width=10cm]{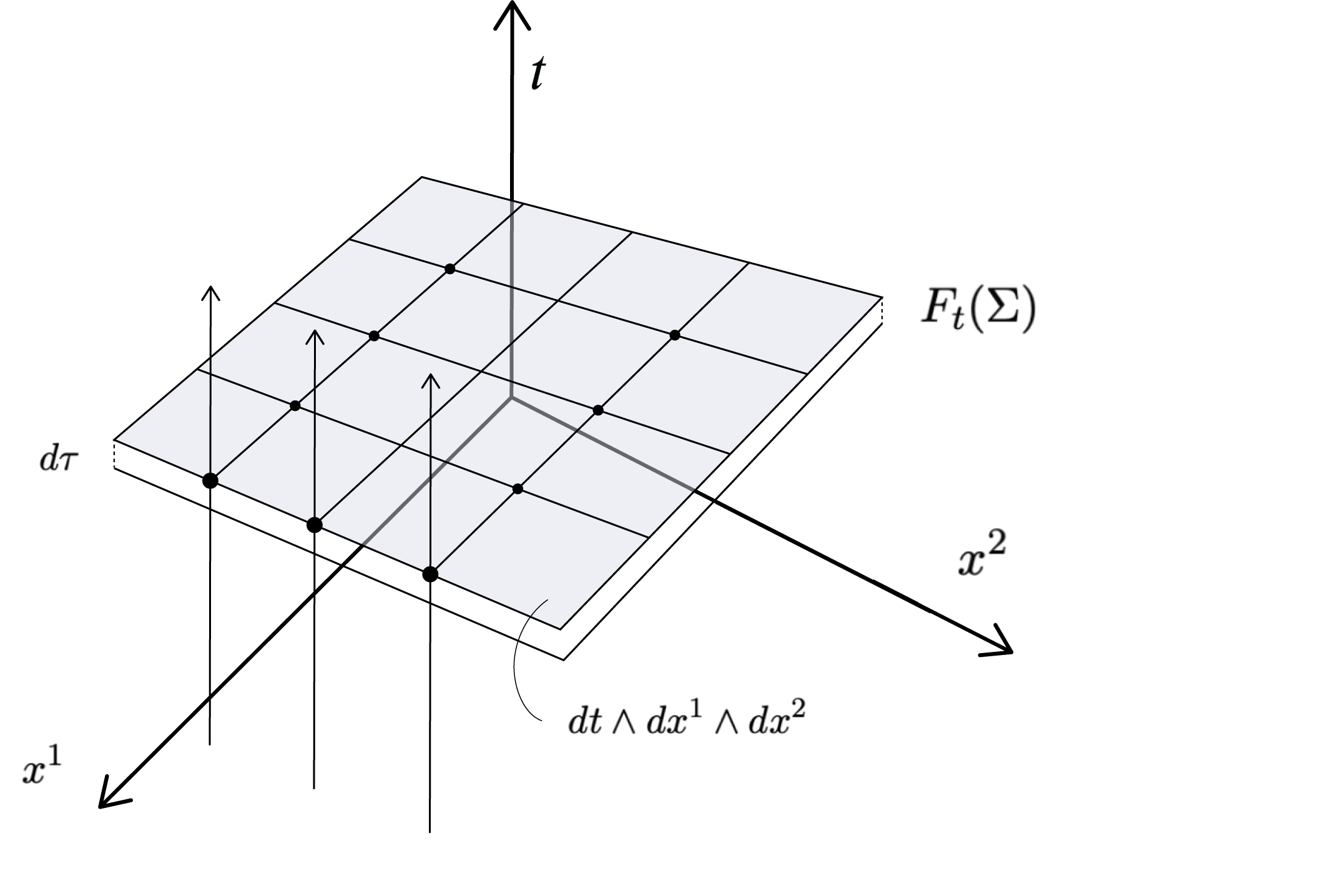}
        \caption{Sketch of integrand $dt\wedge dx^1 \wedge dx^2$,}
    \end{subfigure}%
    ~ 
    \begin{subfigure}[t]{0.5\textwidth}
        \centering
        \includegraphics[width=10cm]{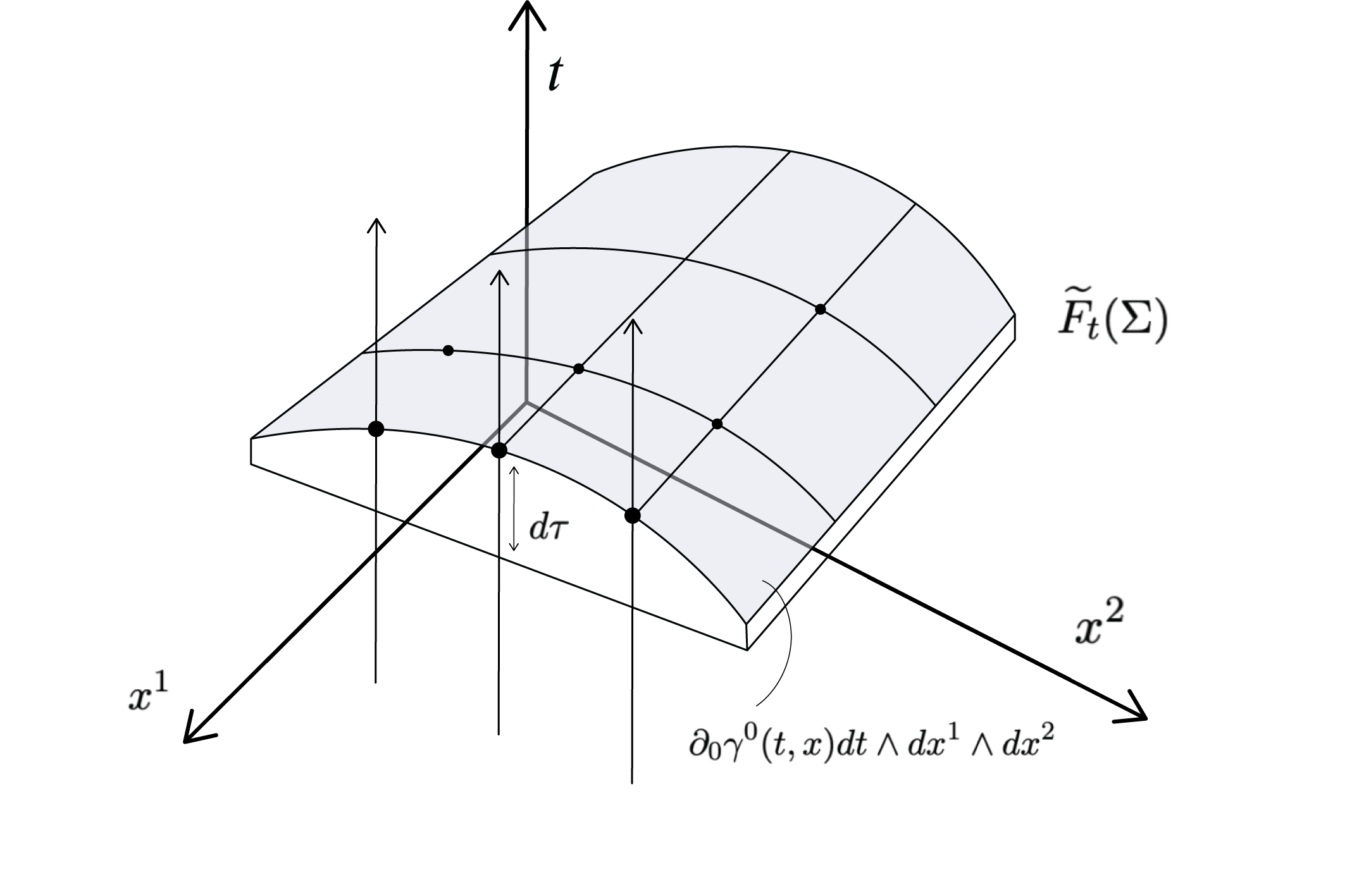}
        \caption{Sketch of integrand $d\tau \wedge dx^1 \wedge dx^2$.}
    \end{subfigure}
        \caption{Depiction of the integrand transformation $dt\wedge dx^1 \wedge dx^2 \mapsto \partial_t \gamma^0 (t,x) dt \wedge dx^1 \wedge dx^2$ in two spatial dimensions. The 2-form $dx^1 \wedge dx^2$ is portrayed here by the squares of intersecting coordinate lines on the spatial hyperplane $F_t(\Sigma)$. The integral under the foliation-GT can intuitively be understood as the transformed 3-volume.}
        \label{fig:AvgGaugedependenceTimecomponent}
    \end{figure*}
    
    \begin{equation}\label{eq:spatially-induced-avg-coord}
        \int_{\varphi_y ( F_t (\D) )} S(t,y) \sqrt{h(t,y)}  \; d^3 y \longmapsto 
        \int_{ \varphi_y ( F_t(\D) )} \gamma^* \big( \partial_t \gamma^0 (t,y) \; S(t,y) \sqrt{h(t,y)} \big) \; d^3 y.
    \end{equation}
    as $dt ( \gamma _* \alpha \n) = \partial_t \gamma^0 (t,y)$. The gauge-dependency thus manifests as the change in time-elapsing rate with respect to the projection along the transformed spacelike hypersurface. Intuitively, it shows how gauge transformation $\gamma$ distorts the time-evolution in the rest-frame, $\partial_t \gamma^0 (t,x) \neq 1$, along every  which is intrinsic to the spacetime-slicing transformation. Note that this is a purely relativistic effect of the gauge transformation. The gauge-dependence is portrayed by a sketch in Figure \ref{fig:AvgGaugedependenceTimecomponent} of how the integrand transforms intuitively.
    
    The total effect of $\gamma$ is seen by incorporating the change of integration domain into the integrand,
    \begin{align}\label{eq:spatial-foliation-GT}
        \int_{F_t(\D)} \gamma^* \iota_{\gamma_* \n} \S 
        &= \int_{\varphi_y (F_t(\D))} \frac{\det_0 (D\gamma)}{\det(D\gamma)} \; \partial_t \gamma^0 (t,y) \; S(t,y) \sqrt{h(t,y)} \;  d^3 y,
    \end{align}
    with $\det_0(D\gamma)$ the determinant of the principal minor submatrix retrieved from $D \gamma$ by removing the first row and column. This term reveals the nature of the spatial 3-form transforming in a three-dimensional way, but its metric determinant is also sensitive to its local time deformation. This points to the implicit notion that it is still the full metric determinant $\sqrt{-g}$ under which the spatial metric is living as its inducement $\sqrt{h}$ is dependent on the foliation $F$ over whole of $M$. This dependence thus is crucial in the gauge transformation of the spatial 3-average.

\subsection{Gauge-invariant averaging: the Gasperini suggestion}
    The implication of our discussion, understanding gauges as foliation choices, leads to an elegant clarification of the intricacies that could arise from a particular specification of averaging. As an example, we consider an averaging method that has been introduced before; it has been suggested in order to solve the gauge dependence problem, see e.g. Marozzi \cite{marozzi2011backreaction}.
      
    Scalar averaging $\langle \cdot \rangle^{(3)}$ is not invariant under the diffeomorphisms on $M$. To overcome this problem, Gasperini et al. \cite{gasperini2009gauge, gasperini2010covariant, gasperini2011lightcone} construct and utilize an averaging procedure,
    \begin{equation}\label{eq:avg-procedure-gasperini}
        \langle S \rangle^\circ_\mathcal{D} (\tau) := \frac{ \int_{\mathcal{D}(\tau)}   \widehat{S}(\tau,x) \sqrt{\widehat{h} (\tau,x) } \; d^3 x  }{ \int_{\mathcal{D}(\tau)} \sqrt{\widehat{h}(\tau,x) } \; d^3 x  },
    \end{equation}
    where $\widehat S = S \circ \phi^{-1}$ and $\widehat h = h \circ \phi^{-1}$ with $\phi$ being the map $\phi(t, x) = \big(\phi^1(t,x), x \big) = (\tau, x)$ such that the foliation are now slices over $\phi(t, x) =\text{const}$.

    Considering the auxiliary map $\phi$ makes visible that $\langle \cdot \rangle^\circ$ weighs the contributions of $S$ in a particular way. The averaging procedure is constructed in such a way that it weighs the scalar field contributions over the slice of the transformed foliation $\phi(\Sigma)$, where we understand $\phi$ as a foliation change $\phi: \Sigma \to \phi(\Sigma)$. Notice the subtle point here: we are not weighing $S$ with respect to its implicitly fixed foliation $\Sigma$ but $\phi(\Sigma)$. This is problematic as $S$ can physically only be considered with respect to $\Sigma$, and not $\phi(\Sigma)$.
    
\begin{figure}[ht]
        \centering
        \includegraphics[width=15cm]{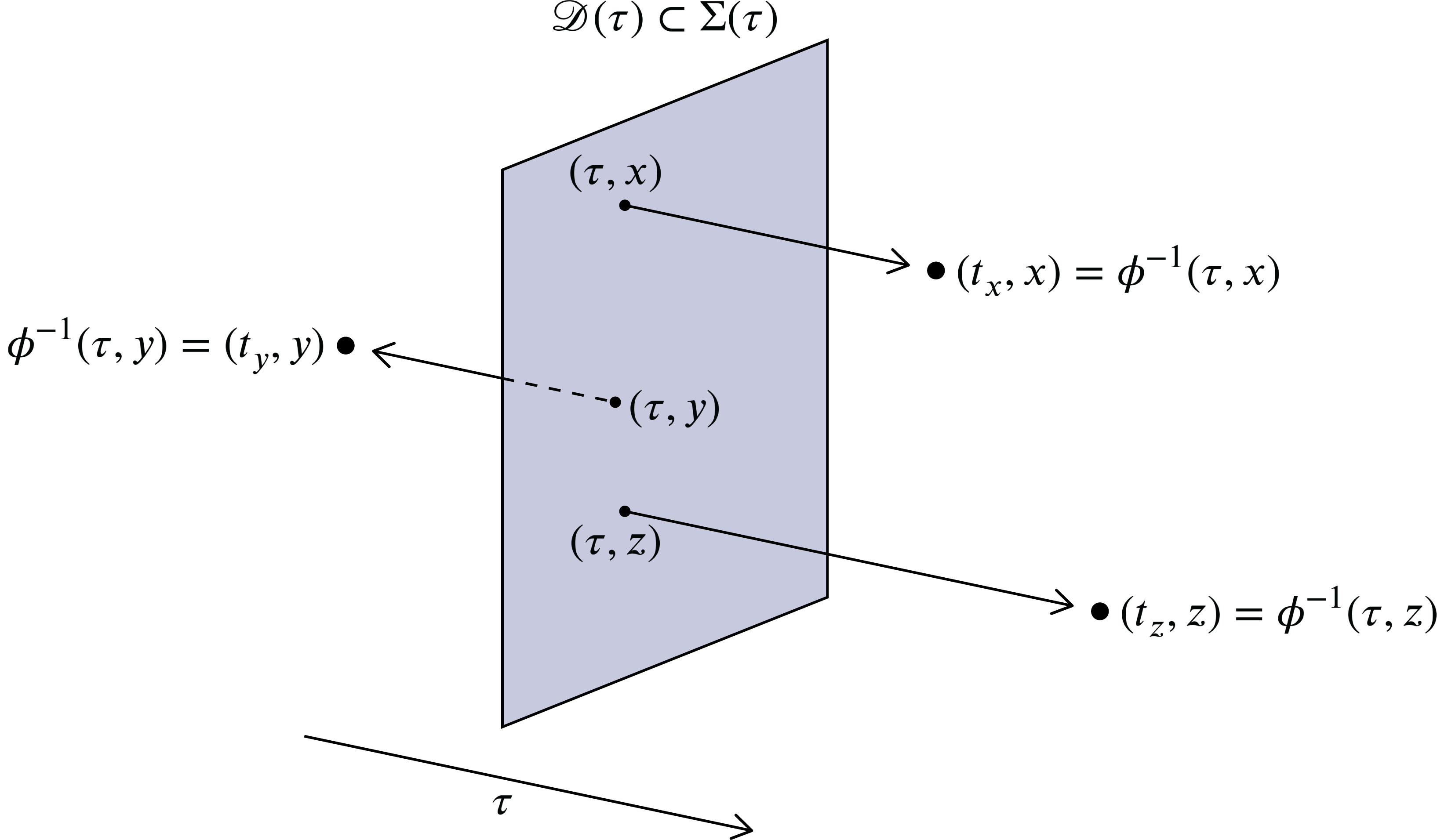}
        \caption{Sketch of foliation change $\phi^{-1}$ for $\mathcal{D}(\tau)$ in slice $\Sigma(\tau)$.}
        \label{fig:nonequivalent-averages-transformation}
    \end{figure}

    To see this, the foliation has been specified to be the family $\big\{ \Sigma( \phi^1(t,x) ) \big\}$, where one slice is of the form
    \begin{equation}
        \Sigma( \phi^1(t,x) ) = \big\{ (t,x) \in M \mid \tau = \phi^1(t,x)  \big\}.
    \end{equation}
    Then we can rewrite \eqref{eq:avg-procedure-gasperini} in terms of the fields $S$ and $h$ as
    \begin{equation}\label{eq:avg-procedure-gasperini2}
        \langle S \rangle^\circ_\mathcal{D} (\tau) = \frac{ \int_{\mathcal{D}(\tau)} S(t_x, x) \sqrt{h (t_x, x) } \; d^3 x }{ \int_{\mathcal{D}(\tau)} \sqrt{h (t_x, x) } \; d^3 x },
    \end{equation}
    where $t_x$ is the time-coordinate value such that $\phi^1(t_x, x) = \tau$. We explicitly invoke subscript notation $t_x$ to indicate that when integrating over $\phi(t_x,x) = (\tau, x) \in \mathcal{D}(\tau)$ in \eqref{eq:avg-procedure-gasperini2}, the value of the time-coordinate $t_x$ varies and is thus \textit{not} constant in $D(\tau)$. Comparing $\langle \cdot \rangle^\circ$  written in the form above with $\langle S \rangle^{(3)}$, we conclude that they are fundamentally different as they average over different points in the spacetime: $(\tau, x) \neq (t_x, x)$ as generally $\phi^1(t_x, x) \neq t_x$. Figure \ref{fig:nonequivalent-averages-transformation} intuitively depicts the transformation $\phi$.

    Thus, when comparing $\langle S \rangle^\circ_{\mathcal{D}} (\tau)$ to $\langle S \rangle^{(3)}_{\mathcal{D}} (\tau)$, the integration region $\mathcal{D}(\tau)$ is the same, but the scalar fields $S$ and $\sqrt{h}$ under the integral are evaluated at different spacetime points. Hence, generally, for a scalar field $S:M \to \mathbb{R}$
    \begin{equation}
        \langle S \rangle^\circ_\mathcal{D} (\tau) \neq \langle S \rangle^{(3)}_\mathcal{D} (\tau) .
    \end{equation}
    This, furthermore, proves that these two averaging procedures are not invariant under foliation-gauge transformations, and thus \eqref{eq:avg-procedure-gasperini} as well. 
    
   From the perspective of the integration domain, we do not weigh the field contributions in $\mathcal{D}$, but the transformed spacetime points ending up in $\mathcal{D}$ by the transformation $\phi$. This is counter-intuitive to what the foliation is in the first place: to fix the spacetime points we consider as `spatial' from which we induce the spatial region $\mathcal{D}$. This supports the mathematical reality that $\widehat{S}$ cannot be identified with $\mathcal{D}$, as the averaging construction implicitly induced the cosmological equivalence class such that $S$ is identified with $\mathcal{D}$.
   
    Formally, we pick beforehand a representation $(M, g, \{F_i\})$ from its equivalence class. As was proven above, there is a one-to-one correspondence between $M$ and the foliation $\Sigma$. Thus, the scalar field $S \in \{F_i\}$ is identified with $\Sigma$. In the context of the foliation-gauge transformation $\phi$ considered above, the cosmology representation is mapped
    \begin{equation}
        \big( \Sigma, g, \{ F_i \} \big) \longmapsto \big( \phi(\Sigma), \phi_* g, \phi_* F_i \big) .
    \end{equation}
    Specifically, $\phi_* S$ is identified with $\phi(\Sigma)$ and not with $S$ itself. Averaging $S$ over slices of the foliation $\phi(\Sigma)$ is thus unambiguous.

    To address whether a gauge-invariant averaging can be constructed which is suitable for cosmic structure analysis, we turn to the theory of gauge-invariant perturbations.

\subsection{Spatially inducing along average particle flow}\label{subsec:InducingAlongAverageParticleFlow}

    Here we briefly focus on the mathematical relation of the particle flow, the foliation it induces, and the relation to the averaging in backreaction studies. We address this in line with smooth manifold theory.

    An arbitrary nonzero timelike vector field $\u:M \to TM$ on spacetime $M$ has a spacelike tangent surface $D$ orthogonal to the tangent spaces of each of the maximal connected integral curves of $\u$. Here $D$ is a linear subspace, also called a \textit{distribution}. The corresponding hypersurfaces $\{\F_i\}$ orthogonal to the integral curves are the embedded submanifolds of $M$ that have $D \subset TM$ as tangent space, i.e. $T_p \F_i = D_p$ for every $p \in \F_i$, and form a foliation $\bigsqcup_i \F_i \cong M$ if any of the following equivalent conditions hold:
    \begin{enumerate}
         \item[(a)] The distribution $D \subset TM$ is involutive, that is, $[v_1, v_2]_p \in D_p$ for all $v_1, v_2 \in D_p$.
        \item[(b)] For the natural 1-form $\nu := g(\u,\cdot)$ we have $\nu \wedge d \nu =0$ everywhere on $M$.
        \item[(c)] The vorticity of $\u$ vanishes, cf. \cite[Ch.~6]{stephani2003exactsolutions}.
        \item[(d)] There is no shell-crossing of the streamlines\footnote{Streamlines are the integral flow lines of a vector field representing the averaged flow of fluid particles.}, cf. \cite{Rampf2021shellcrossing, ShethWeygaert2004MuchAdoAboutNothing}.
    \end{enumerate}
    The first statement is a direct result due to the Frobenius theorem for distributions. 

    The choice of nonzero timelike vector field $\u$ satisfying the above conditions, and thus the induced foliation, remains rather arbitrary. If a (global) temporal function $\t: M \to \mathbb{R}$ exists and is given, one can define a naturally corresponding $\u := -\alpha \nabla \t$ as in \eqref{eq:foliation-GT}. Its construction is however a matter of choice. Ideally, one likes to think of $\u$ as some physical characteristic of the fluid particles on $M$. Its integral curves are then regarded as the paths the fluid particles trace if they flow along vector field $\u$, and thus $\u$ is interpreted as the averaged-out velocity field of the fluid. 
    
    This choice of `averaged' particle flow is scale-dependent and not unique, albeit must be based upon some cosmic observational notion.\footnote{It is also therefore that in from a physics perspective, one might prefer to call certain choices of foliations not to be `gauges' as mentioned by e.g. Buchert \& Räsänen \cite{buchert2012backreaction}. Although the formal usage of gauges in mathematical GR and relativistic cosmology must be pointed out as done in the present work.} The spatial averages can be taken to correspond to observables in a sensible way. Such efforts have been initiated by Räsänen \cite{rasanen2009LightPropogationI, rasanen2010LightPropogationII} and Koksbang \cite{koksbang2019, koksbang2020observations, koksbang2021SignalsOfInhomogeneity}. The freedom of considering an averaged particle flow, and the corresponding gauge transformation between two choices with respect to the induced spatial average, is explicitly seen in \eqref{eq:spatially-induced-avg}.



\section{Case study: An isolated cosmic void}\label{sec:case-study}
On the largest scales, the matter in the Universe forms a complex and tangled structure known as the \textit{cosmic web} \cite{zeldovich1970,joeveer1978,bond1996nature,weybond2008}. This web features clusters of galaxies as nodes connected by filaments and sheets, creating the largest nonlinear structures in the Universe. Its volume is dominated by nearly empty regions called \textit{voids}
\cite{weybond2008b,aragon2010multiscale,weygaert2011,cautun2014}. Their dynamics is primarily influenced by the expansion of the universe, local excess expansion induced by the void, and as well by the external gravitational influence by surrounding filaments and walls \citep{icke1984,WeygaertKampen1993MNRASVoidsGravInstab,shethwey2004,weygaert2016,kugul2024}. This represents an ideal setting for exploring the relevance of spacetime foliations in a general cosmological context, and of how they lead to contradictory notions of averaged cosmological quantities. This discrepancy may introduce artificial backreaction terms in the commonly used spatial averaging formalism. We present a heuristic, simplistic model of a cosmic void and compare \textit{inferred} backreaction results from averages over different appropriate foliations.

We focus on a void configuration with the intention to infer any cosmic features, but as a fitting case study. Given the
dynamically largely confined gravitational influence region, they are ideal testbeds to demonstrate that the current averaging framework
has a significant drawback due to its foliation-gauge dependency. We are particularly interested in the local expansion behavior in
the interior and surroundings of voids, where we see the manifestation of both the void driven acceleration as well as the kinematic
effects of backreaction.

To first order, the nonlinear expansion behavior of voids is fully determined by the global cosmic Hubble
expansion and the local excess expansion proportional to the (nonlinear) density contrast \citep{WeygaertKampen1993MNRASVoidsGravInstab,hamaus2014}. This assumes that we may reasonably ignore external influences on the dynamics of the void, although this is only true for the
excess expansion of the void. For the overall dynamics of the void, a recent study has demonstrated that in fact there is a
substantial gravitational influence exerted by the surrounding overdense filaments \citep{kugul2024}. In summary, we opt for
a configuration of an isolated void to model the interior dynamics of the void, augmented by the presence of a filamentary
structure outside its realm. In effect, it is an idealized representation for an uncompensated void model. It also serves
better our purpose, since a requirement to guarantee a nonvanishing kinematical backreaction is that the
configuration is not spherically symmetric \cite{buchert2000cosmparamaters}.

An essential aspect for our assessment is that of the multiscale structure of voids, a direct manifestation of the
hierarchical evolution of the void population \citep{shethwey2004,weybond2008b,aragon2013,weygaert2016,jaber2024,kugul2024}. It translates 
into a related multiscale structure of the corresponding void velocity (out)flow. It is visible as a variation in the local excess
expansion rate \citep{WeygaertKampen1993MNRASVoidsGravInstab,aragon2013,kugul2024}, and has recently also been detected in the observed velocity flows, from the Cosmicflows-3 survey \citep{tully2016}, in and around a sample of voids in the SDSS survey \citep{courtois2023}.

To first approximation, a mature void region evolves into a region of near constant underdensity - with a value of
approximately $\Delta \approx -0.8$ - surrounded by an overdense boundary \citep[see e.g.][]{shethwey2004,weygaert2016}. This
\textit{bucket shape} mass profile corresponds - simply on the basis of the continuity equation - to a superhubble interior expansion
because of the implied constant divergence of the flow field. It would imply its macroscopic evolution to be exactly equal to
its microscopic propreties. Any variation in mass density due to the internal residual multiscale structure of the void will imply deviations from this uniform excess void expansion, and will induce variations in the local excess expansion \citep{WeygaertKampen1993MNRASVoidsGravInstab,aragon2013,kugul2024}. For our purpose, this is highly relevant as these local deviations in expansion from the macroscopic average will appear in the kinematical backreaction term \cite{buchert2000averagedust}.

\begin{figure}[ht]
    \centering
    \includegraphics[width=15cm]{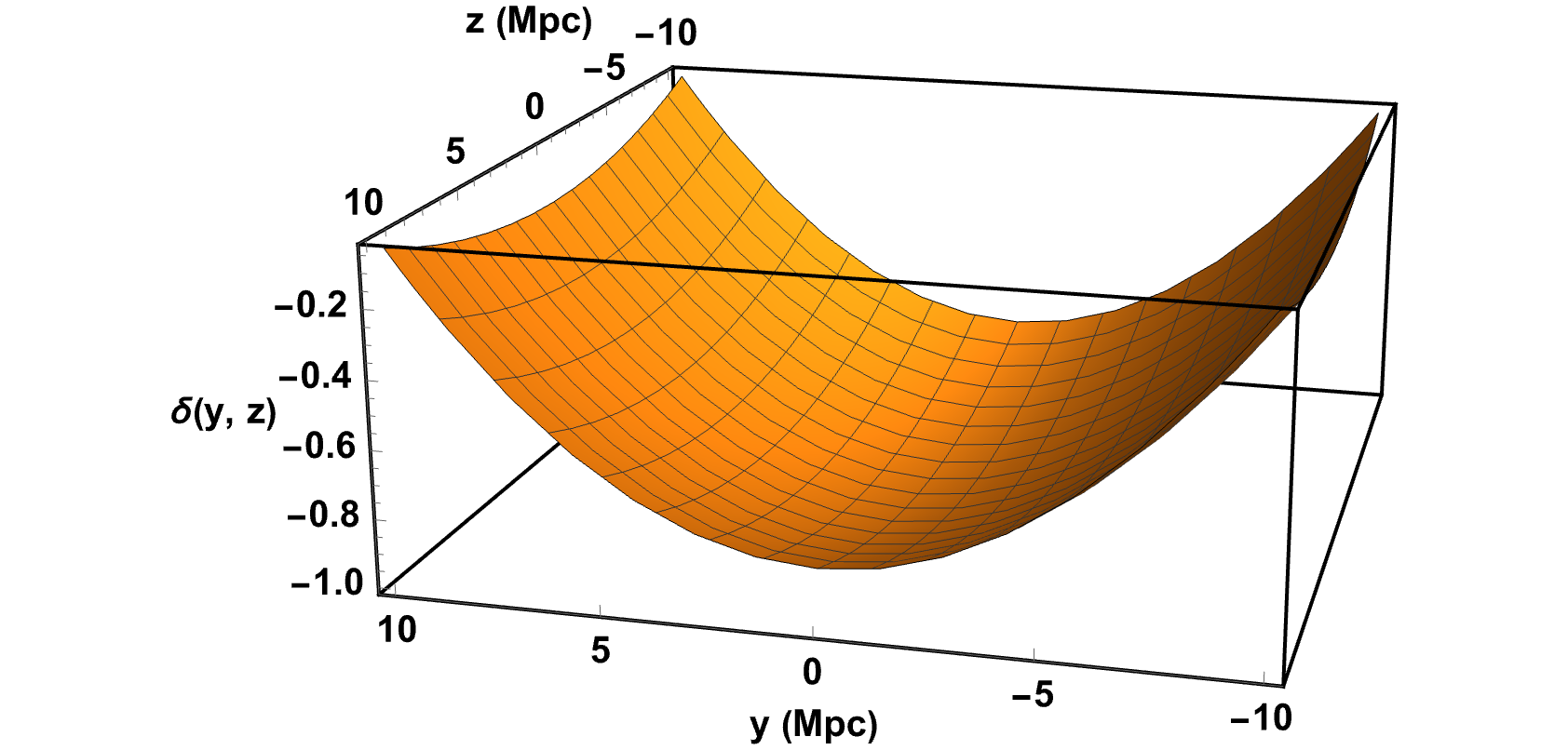}
    \caption{Graph of the density contrast $\delta(y,z)$ as a toy model of an elongated void.}
    \label{fig:void-density}
\end{figure}

\subsection{2D heuristic void model}
\label{sec:voidmodel}
Following the reasoning outline above, we construct a two-dimensional heuristic test model of a void $\D$ embedded in a $\Lambda$CDM universe. To this end, we relax the spherical symmetry of the void and post a neighbouring filamentary feature
at its boundary. We restrict ourselves to a situation in which matter can be assumed to be pressureless, which for most relevant cosmological situations is adequate given the major share of the cosmic matter content is dark matter. Also, for simplicity, we assume the density contrast $\delta$ to remain constant in time. Once voids have become mature troughs at $\delta_{\rm void} \approx -0.8$, this is a
reasonable assumption \citep{blumenthal1992}, although at earlier times of its evolution its density contrast is rapidly
diminishing as mass streams outward.

The mass deficit in the void induces an effective push from the interior to the boundary, leading to mass streaming from the void into the filament. To first approximation, in the conventional Newtonian approximation a purely void spherical would reveal a superHubble void expansion \cite{icke1984,WeygaertKampen1993MNRASVoidsGravInstab,shethwey2004,hamaus2014}, in which the void experiences an excess expansion rate $H_{void}$,
\begin{equation}
  H_{void}\,=\,\frac{\dot a_{\rm void}}{a_{\rm void}}\,=\,\Bigl[ 1+\frac{1}{3}\,f(\Omega_m) \delta_{\rm void} \Bigr]\,H\,\approx\,1.14\,H\,,
\end{equation}
for a void in a $\Lambda$CDM cosmological background, with $f(\Omega_m)\approx \Omega_m^{0.55}$ the structure growth rate. In other words, we may expect a mature void to be expanding with an approximately $14\%$ higher rate. 

\bigskip
For our case study void configuration, we enforce the combination of a void with a neighbouring filament by by imposing an
anisotropic density profile whose contrast grows more rapidly along the $y$-direction than
in the $z$-direction. The model void has a radius of approximately 15 Mpc in the $y$-direction, and 25 Mpc in the $z$-direction. The elongated filaments at its boundaries also have a lengths of $\approx 20$ Mpc. The heuristic expression that we use for modelling this configuration is specified by the density (contrast) profile
\begin{equation}
  \delta(y,z) = -1 + 5 (y/25)^2 + (z/25)^2, \qquad y,z \in [-10, 10] \; \text{Mpc},,
  \label{eq:voidprof}
\end{equation}
with the density contrast $\delta$ related to the density $\rho$ via the usual relation in terms of the global (current epoch)
FLRW density $\rho_{0}$,
\begin{equation}
  \rho(y,z)\,=\,\rho_{0}(1+\delta)\,.
  \end{equation}
Figure~\ref{fig:void-density} shows the run of the density in and around the void. The interior of the void, where it reaches
its minimum depth, is the realm $y \in [-5, 5]$. The boundaries of the void, with the neighbouring filaments, are located at $y=-10$ and $y=10$ and extends along the $z$-coordinate.

\bigskip
For the two-dimensional void configuration that we consider in our case study, we assess spatial averaging of cosmological
quantities $S(t,y,z)$ according to the 2D integral
\begin{equation}\label{eq:void:avg-method}
    \langle S \rangle := \frac{1}{V_\D} \int_\D S(t,y,z) \sqrt{h(t,y,z)} \; dy dz \,.
\end{equation}

\subsection{Averaging and acceleration}
To follow the evolution of the mass distribution in and around the void, we consider the variation of the cosmic expansion
factor $a(t)$ around the global expansion factor $a_{LCDM}(t)$. The background $\Lambda$CDM universe involves the following cosmological parameters: $H_0 \approx 70$ km/s/Mpc, $\Omega_m \approx 0.3$, $\rho_0 \approx 10^{11} \; M_\odot$/Mpc${}^3$ and $G \approx 4 \cdot 10^{-9}$ Mpc km${}^2$ s${}^{-2}$ $M_\odot^{-1}$. 
Given the fact that at the current cosmological epoch, we are rapidly approaching the late-time fully dark energy dominated era, the
asymptotic limit of the de Sitter expansion is a reasonable approximation for the cosmic expansion,
\begin{equation}
a_{LCDM}(t) \approx \exp(t H_0)\,.
  \end{equation}

\bigskip
\noindent The intention of the current case study is to determine the implied \textit{average} expansion factor $a_{\cal D}$
\begin{equation}
  a_{\cal D}\,\equiv\,\langle a \rangle 
  \end{equation}
within a certain region ${\cal D}$, along with the corresponding effective expansion rate $H_{\cal D}$, 
\begin{equation}
  H_{\D}\,=\,\frac{{\dot a}_\D}{a_\D}\,=\,
  \frac{1}{3} \langle \theta \rangle\,.
  \end{equation}
The latter is effectively the average divergence $\theta_{\D}$ of the flow field ${\u}$ within region ${\cal D}$,
\begin{equation}
  \theta_{\D}\,=\,\langle \theta \rangle \,=\,\langle \nabla \cdot {\u} \rangle\,.
\end{equation}
Evidently, when the region $\cal D$ attains a size comparable to the Hubble radius, ie.  $\D \to \Sigma$, the divergence
reaches asymptotically the global value for the Hubble parameter, $\langle \theta \rangle_\Sigma = 3 H_0$.

\bigskip
\noindent To follow the expansion of the void region, we start from the local Friedmann-Lemaitre acceleration equation for
the averaged acceleration $\ddot{a}_\D$ in a patch ${\cal D}$.
\begin{equation}\label{eq:void:AvgAccEq}
  \frac{\ddot{a}_\D}{a_\D} = \Lambda + 4 \pi G \langle \rho \rangle - \frac{2}{9} (1 - \Omega_{k_{\cal D}}) \langle \theta \rangle^2\,,
  \end{equation}
in which $k_D$ is the effective curvature parameter for the patch ${\cal D}$. Corresponding to $k_D$, the effective curvature
density parameter $\Omega_{k_{\cal D}}$ in region ${\cal D}$ --- in units of the critical density $\rho_c$ --- is defined as 
  \begin{equation}
    \Omega_{k_\D}\,=\,-\frac{k_{\D}\, c^2}{a_{\D}^2 H_\D^2\,R_0^2}.
    \end{equation}
      with $R_0$ the current curvature radius of the Universe. 
  Following the expression inferred by Buchert \cite{buchert2000averagedust}, the averaged acceleration $\ddot{a}_\D$ of the
  local cosmic expansion is given by 
\begin{equation}
    \frac{\ddot{a}_\D}{a_\D}
    = \frac{1}{3} \big( \Lambda - 4 \pi G \langle \rho \rangle +  Q_\D \big), 
\end{equation}
in which $Q_\D$ is the kinematical backreaction term. Its value is given by 
\begin{equation}\label{eq:void-kinematicbackreaction}
    Q_\D = 2 \Lambda + 16 \pi G \langle \rho \rangle - \langle  {{}^3 \cal R} \rangle  - \frac{2}{3} \langle  \theta \rangle^2,
\end{equation}
in which $\langle \rho \rangle$ is the averaged density over the void path ${\cal D}$ and ${{}^3 \cal R}$ the spatial Ricci
scalar on a spatial hypersurface \footnote{Note that this concerns the spatial Ricci scalar ${{}^3 \cal R}$, which is derived from the 3-dimensional Ricci tensor ${{}^3 \cal R_{ab}}$ of the spatial hypersurface $\Sigma_t$. This is different from the spacetime Ricci scalar ${{}^4 \cal R}$, which is derived from the 4-dimensional spacetime Ricci tensor  ${{}^4 \cal R_{ab}}$ of the full spacetime $M$.}. The spatial Ricci scalar over the volume ${\cal D}$ is a function of its local curvature $k_{\D}$
and local averaged expansion rate $ \theta_\D$,
\begin{equation}
  \langle {{}^3 \cal R}\rangle \,=\,- \frac{2}{3} \langle \theta \rangle^2 \Omega_{k_\D}\,.
\end{equation}
Also note that when the patch $\cal D$ becomes comparable to the global universe we have $Q_\D \to 0$ and $\Omega_{k_\D} \to 0$, so
that the acceleration equation becomes asymptotically equal to the global Friedmann-Lemaitre equation. 

\bigskip
\noindent Following the formalism outlined in the previous sections, the average acceleration $a_{\D}=\langle a \rangle$ in and around
the cosmic void patch in a $\Lambda$CDM background model is given by
\begin{equation}\label{eq:void:acc-aprox}
  \frac{\ddot{a}_\D}{a_\D}
  = H_0^2 \bigg( 1 + 2 \Omega_{k_\D} + 4 \pi G \rho_0 H_0^{-2} (1 + \langle \delta \rangle) - (1- \Omega_{k_\D}) \big( \tfrac{2}{9} \Omega_m^{1.2} \langle \delta \rangle^2 - \tfrac{4}{3} \Omega_m^{0.6} \langle \delta \rangle  \big)  \bigg)\,.
\end{equation}
Given at the current epoch the dark energy dominated universe is asymptotically approaching the de Sitter exponential expansion
phase, for simplification we use its approximate relationship between cosmological constant $\Lambda$ and Hubble parameter $H_0$,  
$\Lambda = 3H_0^2$. To obtain the relation above, we use the first order approximation for
the implied excess expansion rate \citep{Lahav1991MNRAScosmologicalconstant, WeygaertKampen1993MNRASVoidsGravInstab}
\begin{equation}
  \langle \theta \rangle = 3H_0 - \Omega_{m,0}^{0.6} H_0 \langle \delta \rangle.
  \end{equation}

\bigskip
\noindent For the specific case of the two-dimensional heuristic void model outlined above \eqref{eq:voidprof}, we translate the nearly constant internal density profile at $\delta \approx -0.8$
(see e.g. \cite{blumenthal1992,WeygaertKampen1993MNRASVoidsGravInstab,shethwey2004}) to an internal curvature density parameter $\Omega_{k_\D}=0.2$ throughout the void (also see recent relativistic numerical work on kinematic backreaction by Williams et al. \cite{Williams2024arXiv:VoidStatisticsNumericalRelativity}). By integrating the acceleration equation
for the void patch \eqref{eq:void:acc-aprox}, we obtain the following approximate solution for the average scale factor $\langle a\rangle_F (t)$ in and around the void 
\begin{equation}\label{eq:void-avg-scale-factor}
    \langle a\rangle_F (t) \approx \exp t H_0 \bigg( 1.4 + 4 \pi G \rho_0 H_0^{-2} \langle 1 + \delta \rangle_F   - 0.8 \big( \tfrac{2}{9} \Omega_m^{1.2} \langle \delta \rangle_F^2 - \tfrac{4}{3} \Omega_m^{0.6} \langle \delta \rangle_F  \big)  \bigg),
\end{equation}
in which $F$ indicates the dependence on the specified foliation (see Section~\ref{sec:foliation}). 

\subsection{Foliation}
\label{sec:foliation}

With respect to the averaged scale factor \eqref{eq:void-avg-scale-factor} we denote the foliation dependence explicitly.  The approach is to consider different foliations and see how it changes the numerical value of $\langle a \rangle_F$ for our simple void model. 

Let us consider the unperturbed conformal Newtonian foliation-gauge $F_{conf}$ as discussed in Section~\ref{subsec:relation-between-foliations-and-perturbation-theory}, that is,
\begin{equation}
    F_{conf} \quad s.t. \quad \alpha = 1, \qquad \beta_i = 0, \qquad h_{ij} = \delta_{ij}.
\end{equation}
We gauge-transform $F_{conf}$ to generate other foliations $\widetilde{F}$. Recall from Section~\ref{subsec:averaging-dependence-on-foliation-choice} that for both of the foliations, even in the general 3-dimensional case, we can retrieve the standard spatial form \eqref{eq:void:avg-method} by writing
\begin{equation}\label{eq:void-avg-transformation}
    \langle S \rangle_{F_{conf}} 
    = \frac{1}{V_\D} \int_{\D} S(t,x) \sqrt{h(t,x)} \; d^3 x;
    \qquad 
    \langle S \rangle_{\widetilde{F}} 
    = \frac{1}{V_\D} \int_{\D} S(t,\tilde{x}) \sqrt{h(t,\tilde{x})} \; d^3 \tilde{x} ,
\end{equation}
in coordinates adapted to the corresponding foliations. Here we suppressed the foliation dependence in the integration domain as commonly done in backreaction studies, e.g. \cite{buchert2000averagedust}. Specifically, we consider two different foliation-gauge transformations $\gamma$ generating different foliations. We consider an infinitesimal diffeomorphism $\gamma_{inf}$ mapping $x^\mu \mapsto \gamma_{inf}^\mu (x) = x^\mu + \xi^\mu$ that generates the foliation $F_{inf}$, and a gauge transformation $\gamma_{flrw}$ that is such that $\dot{\gamma}_{flrw}^0  \approx \widetilde{\alpha} = H_0 /\sqrt{\rho} \propto 1 / \sqrt{1+ \delta}$, conform to the result of Alatas et al. \cite{Alatas2020ILapseFunction} capturing the FLRW effect, generating $F_{flrw}$.

We numerically calculate $\langle a \rangle_F$ for $F_{conf}$ and the generated foliations $\widetilde{F} \in \{ F_{inf}, F_{flrw} \}$. As derived in Section~\ref{subsec:averaging-dependence-on-foliation-choice}, we utilize the foliation-gauge transform of the averaging,
\begin{align}
    \langle S \rangle_{F_{conf}} 
    \longmapsto 
    \langle S \rangle_{\widetilde{F}}
    =
    \bigg \langle \frac{\det_0 (D\gamma)}{\det(D\gamma)} \; \dot{\gamma}^0 S \bigg \rangle_{F_{conf}},
\end{align}
For the conformal Newtonian foliation-gauge we have $\sqrt{h} = 1$, and for $F_{inf}$ we can approximate $\sqrt{h_{inf}} \approx 1$ and $\text{det}_0(D\gamma_{inf}) / \det(D\gamma_{inf}) \approx 1$ as ${(D\gamma_{inf})^\mu}_\nu = {\delta^\mu}_\nu + \partial_\nu \xi^\mu$ and $\xi^\mu$ is infinitesimal. Since there is only a condition for $\gamma_{flrw}$ on the time-derivative of its 0-component, we can construct it such that $\sqrt{h_{flrw}} \approx 1$ and for simplicity we set $\text{det}_0(D\gamma_{flrw}) / \det(D\gamma_{flrw}) \approx 1$. For the infinitesimal foliation transformation we numerically set $\dot{\gamma}_{inf,c}^0 := 1 + \dot{\xi}^0 = 1+ c \sqrt{y^2 + z^2}$ with some small constant $|c| < 1$. 

\subsection{Void expansion: results}
\begin{figure}[ht]
    \centering
    \includegraphics[width=17cm]{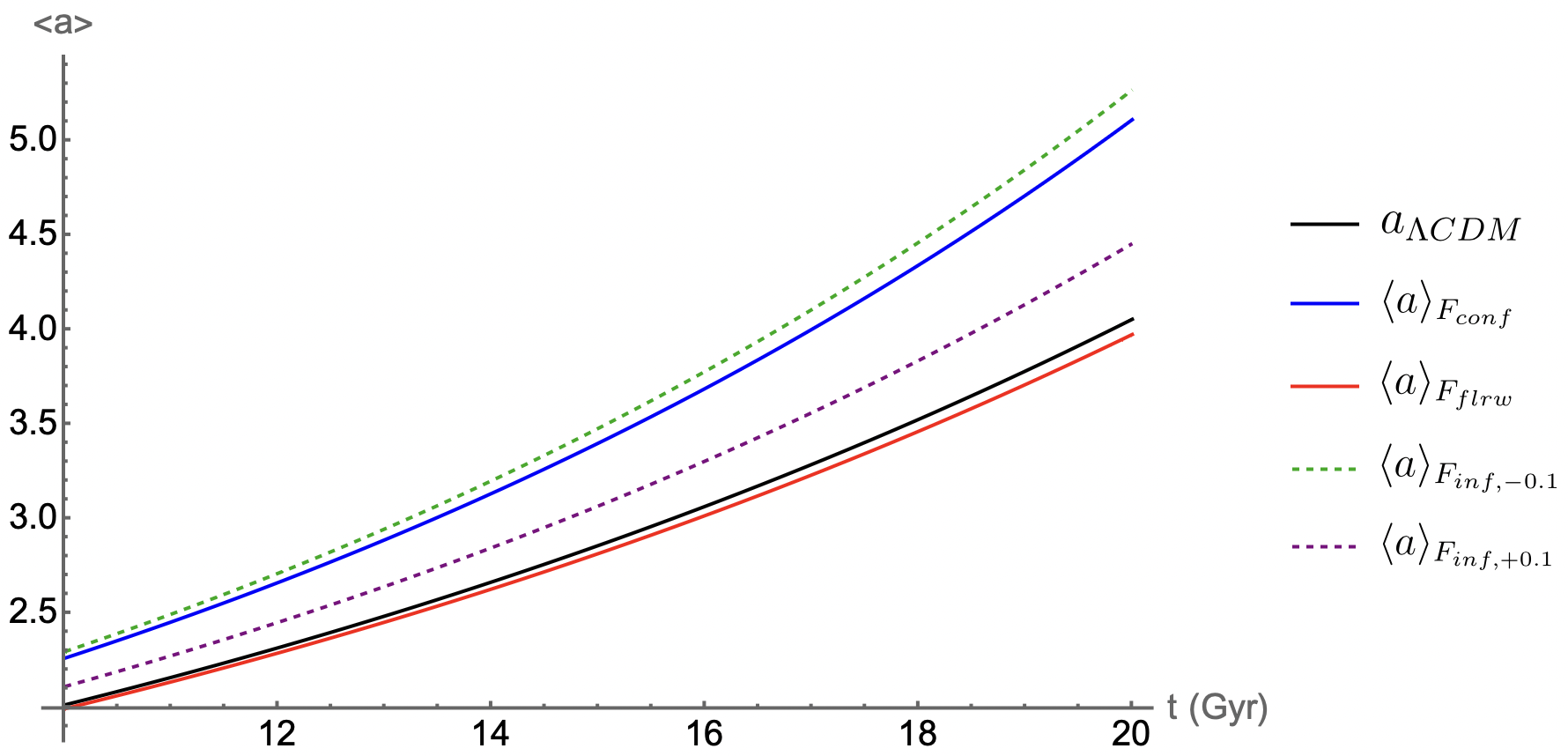}
    \caption{The averaged scale factor $\langle a \rangle_F (t)$ for the different foliations, and global $a_{\Lambda CDM}(t)$ for
      comparison. Here $t=0$ denotes present time.}
    \label{fig:void-scalefactor}
\end{figure}

For the different specified foliations, the numerical solutions of the averaged scale factor $\langle a \rangle_F$ for the specified
void configuration \eqref{eq:voidprof} are depicted in Figure~\ref{fig:void-scalefactor}. We immediately observe
considerable differences between the various options for the foliations, hence reflecting the strong dependence
of foliation on the observed expansion and evolution of a void region.

The black solid line represents the global expansion factor $a_{\Lambda CDM}$ for the background $\Lambda$CDM cosmology. There is a substantial increase of the void expansion from this for the conformal Newtonian foliation-gauge $F_{conf}$ (blue solid line). It
yields an excess expansion that is around 17\% higher than the global background $\Lambda$CDM expansion, comparable to the
expected excess expansion in the Newtonian approximation (see Section~\ref{sec:voidmodel} above).

For the other foliations generated by infinitesimal foliation-transformations, the averaged scale factors (the green and purple dashed lines) show significant deviations from $\langle a \rangle_{F_{conf}}$. The inferred expansion deviates several percentage points for
infinitesimal transformations with $|c| \leq 0.1$. In a few cases, it even implies surprising behaviour entailing a systematic
different nature of the acceleration within the void. This concerns in particular the foliation $F_{flrw}$ (solid red line). In
this particular situation, the expansion of the void would be even less than the global average, by approximately 2\%. Cosmic
observers along this foliation would infer that the local void expands less rapidly than the global expansion of the Universe.

\subsection{Implication: foliation dependence}
In this simple toy model, we see that the foliation dependence for kinematical backreaction term \eqref{eq:void-kinematicbackreaction} is of importance for the inferred results using spatial averages in a relativistic context. It is clearly foliation-gauge dependent, and in that sense artificial. 

This confirms the results found by Adamek et al. \cite{Adamek2019CQG}. They suggest it is therefore not relevant to consider a potential foliation which admits a sizeable backreaction, but the question is whether there exist foliations that have negligible backreaction.

The pivotal message we communicate is the fundamental nature of foliations being gauges, whose freedom is intricately linked to diffeomorphism invariance as elaborated upon in Section \ref{section:foliation-gauges-in-rel-cosmology}. A particular result of our foliation-gauge framework is how the induced spatial averaging is transforming under foliation gauge transformations as for example seen in \eqref{eq:void-avg-transformation}. This raises the question of to what extent one should understand the physical nature of a foliation, and how this understanding influences the inference of backreaction, as investigated by Adamek et al. \cite{Adamek2019CQG}. Buchert et al. \cite{buchert2012backreaction, buchert2018dependencefoliation} have suggested motivating specific choices of foliation-gauges based on their perceived ``physical" nature. In the following Section, we propose a more general approach.




\section{Gauge-invariant spatial averaging for generalized proper time foliations}\label{sec:gauge-invariant-avg-proper-time-foliations}
We have shown that the acclaimed kinematical backreaction term in a relativistic cosmological setting is foliation-gauge dependent. The provided toy model made up that the foliation dependence in averaging quantities can be significant, and altering the inferred nature of the model at hand. Here we provide a practical solution suitable for standard cosmological simulations and relatable to observational work. We identify a subgroup of foliations on which all cosmic observers can agree upon, which we call \textit{generalized proper time foliations}. In line of the above cosmic foliation-gauge framework, we show that the natural spatially induced averaging along the generating flow of the foliation is indeed foliation-gauge invariant for generalized proper time foliations. We comment on the practicality of our solution for cosmological simulations, and leave details for upcoming work.

The cosmic rest frame is the reference frame that comoves with the cosmic microwave background (CMB) radiation over very large scales, and is thus also called the CMB frame \cite{lineweaver1997cmbframe}. The convergence of the large-scale structure to the unique CMB frame is a widely implemented. Although there is some discussion regarding the scales of $\D$ for which the convergence is appropriate, cf. \cite{kraljic2016JCAPuniformHubbleFlow}, the background quantities in the cosmic rest frame are widely used and proven successful. There seems to be at least a consensus and enough proof that there must be some cosmic smoothing scale, say 2 Gpc, to which the cosmic web admits the CMB frame.

The cosmic rest frame is mathematically defined by the corresponding four-velocity $u_{cmb}^a = {\delta^a}_0$ generating foliation $F_{cmb}(\tau)$ with proper time $\tau$. The CMB foliation is equivalent to the proper time foliation of Minkowski background spacetime, and thereby is the natural relativistic generalization of the Newtonian foliation as there is a unique concordance between these foliations, cf. Delphenich \cite{Delphenich2002Foliations-ProperTime}. Here we consider all foliations $F(t)$ that are equal to the cosmic rest frame $F_{cmb}(\tau)$ up to time-reparametrizations and spatial foliation-gauge transformations, that is, 
\begin{equation}\label{eq:ProperRimeAvg-GT}
    F^\mu = \gamma^\mu \circ F_{cmb} \qquad s.t. \qquad \dot{\gamma}^i = 0,
\end{equation}
for all $i=1,2,3$ for any such $\gamma \in \diff(M)$. Let us call this collection of gauge transformation $\mathcal{G}^*$, which we can consider to be generating `generalized proper time' foliations. The physically practical nature of $\mathcal{G}^*$ is worthwhile to consider. We can do this due to the nature of the scale on which the CMB dipole\footnote{Here we assume the kinematic dipole to be of the order $10^{-3}$ and cosmic dipole of the order $10^{-5}$, and thus negligible to the overall CMB dipole.} manifests: it is large-scale, and approximately has constant velocity over time. To generalize this approach, one could account for the small time deviations in the dipole transformation. This should give even better estimates, and is most likely achievable as the changes are small, as then the $\dot{\gamma}^i$ terms are well-approximated by first order terms. This is left as a direction for future research.

Cosmological observers in a perturbed Universe can always measure their peculiar velocities and observe their CMB dipole. Transforming their peculiar local frame into a frame for which the CMB dipole vanishes constitutes a generalized proper time frame $F(t)$ in the equivalence class $[F_{cmb}(t)]$ under $\mathcal{G}^*$, that is there exists a foliation-gauge transformation $\gamma$ such that \eqref{eq:ProperRimeAvg-GT} holds. We disclaim that this counters the idea of relativity of an absolute preferred frame, it is merely that all cosmic observers can agree upon at least one foliation $F$ generated by $\mathcal{G}^*$. Physical descriptions should be independent of the chosen frame, although cosmic averaging has been proven to be generally dependent on them, here we consider a group of frames for which we can gauge-invariantly average.

It must be noted that there are some implicit assumptions made in the above reasoning for cosmic observers to be able to construct such generalized proper time foliation. Let us coarse-grain just enough to yield a high-resolution cosmological model, say 1 kpc, and denote the four-velocity of a set of cosmic observers in the relativistic fluid by $\u$. Measuring the cosmic peculiar velocities and especially the CMB dipole along $\u$ makes that \textit{de facto} we smooth over a larger scale, say 1 Mpc. The model is thus described by the averaged-out four-velocity $\bar{\u}$, and not $\u$. This is of importance since constructing a generalized proper time foliation is not possible for fluid flows with rotation, and thus vorticity; recall Section~\ref{subsec:InducingAlongAverageParticleFlow}, and see \cite{Delphenich2002Foliations-ProperTime}. Surely, the high-resolution and highly perturbed $\u$ admits vorticity, although it is not far fetched to assume $\bar{\u}$ to be irrotational. To what degree the relativistic kinematical backreaction effects are still captured in $\bar{\u}$, and thus generalized proper time foliations, is left for further research.

Here we consider a solution for the foliation-gauge problem of averaging in relativistic cosmology. Take any generalized proper time foliation $F_t, \widetilde{F}_t \in [F_{cmb}(t)]$, then
\begin{equation}\label{eq:ProperRimeAvg-GI}
    \int_{\widetilde{F}_t(\D)} S_t(x') \sqrt{h_t (x')}  d^3 x'
    = \int_{F_t(\D)} \frac{\det_0 (D\gamma)}{\det(D\gamma)} \dot{\gamma}^0 S_t(x) \sqrt{h_t (x)}  d^3 x 
    = \int_{F_t(\D)} S_t (x) \sqrt{h_t (x)} d^3 x,
\end{equation}
as $\dot{\gamma}^i=0$. Hence, spatially induced averaging along the velocity flow $\u$ is invariant under gauge-transformations $\mathcal{G}^*$ of generalized proper time foliations. 

Ideal to our proper time approach is that it lends us to work in the standard $\Lambda$CDM framework of background scalars. Specifically, when considering a scalar $S(t,x)$ we work with a predefined background $\overline{S}(t)$ such that there is a unique perturbation $\delta S= S - \bar{S}$. Implicitly this assumes that for foliation $F_t$ determined based on a local cosmic flow, there is an straightforward global background $\bar{F}_{\tau}$ to which quantities averaged over all of $F_t(\Sigma)$ tend to. Because we can identify its global Hubble flow, which induces a unique foliation $\bar{F}_{\tau}$. In terms of foliations, standard cosmology takes
\begin{equation}\label{eq:cmb-convergence}
    \frac{1}{V_\D} \int_{F_t(\D)} S(t,x) dV \longrightarrow{} 
    \frac{1}{\bar{V}_\Sigma} \int_{\bar{F}_{\tau}(\Sigma)} S(\tau,\bar{x}) d\bar{V}
    \qquad \text{if} \quad \D \to \Sigma,
\end{equation}
where $V_\D =\int_{F_t(\D)} dV$ and $dV = \sqrt{h(t,x)}d^3x$; idem for the integral quantities of the CMB frame.\footnote{Unlike before, here we leave out the explicit coordinate map $\varphi_x$ in the integration domain.} Although there is some discussion regarding the scales of $\D$ for which the convergence is appropriate, cf. \cite{kraljic2016JCAPuniformHubbleFlow}, the background quantities in the cosmic rest frame are widely used and proven successful. Here we utilize the current notion of the CMB frame, as our goal is to contribute and tap it to standard $\Lambda$CDM cosmology. In other words, we understand background $\overline{S}(t)$ to be in observed in the CMB frame. For any generalized proper time foliation $F_t$ we can integrate out the background $\overline{S}$. This gives
\begin{equation}
    \langle S \rangle_F (t) 
    = \frac{1}{V_\D} \int_{F_{cmb}(t, \D)} (\overline{S} + \delta S)(t, x) \sqrt{h(t,x)} \; d^3 x
    = \overline{S}(t) + \langle \delta S \rangle_F(t),
\end{equation}
using the gauge-invariance relation \eqref{eq:ProperRimeAvg-GI}. Note that this is a unique decomposition as we assume background $\overline{S}(t)$ is uniquely \textit{a priori} determined with respect to the CMB frame. This is common practice, e.g. for the Universe's global average mass density $\rho_0$.

This will be, in particular, interesting for the modelling of structure formation and evolution in FLRW cosmologies. Almost all cosmological studies of structure formation follow the standard approach of assuming a spatially independent background for scalar fields. By complementing this with a unique formulation of the scalar deviations such as described above, and potentially even vector and tensor deviations, translates into a standard formalism for quantifying the averaging of such deviations and for the specification of these in local regions of the large scale matter and galaxy distribution, that is, of the cosmic web. It renders the perturbative averaging applicable to the concordance model of cosmology and other relevant FLRW cosmologies, and applicable to standard simulations.

The vast majority of such computer simulations are based on a pure Newtonian treatment of the gravitational interactions. They assume that this forms a reasonable approximation for the limited size of the simulation boxes and the gravitational potential perturbations represented in these volumes. Not only do such computer simulations follow the standard treatment of averaged quantities, they also involve implicit averaging. Most computer simulations represent the mass distribution on a grid, which involve implicit low-pass and high-pass filters dictated by the size of the simulation box and the grid-cells.

Ideally, the effect of the inhomogeneous mass distribution on the expansion of the universe should be followed by a fully relativistic treatment of the implied gravitational interactions. In recent years there
have been several studies seeking to develop a numerical relativistic code --- for ideal fluids see \cite{east2018Nbody_NewtonianGRcomparison, giblin2016Nbody_DeparturesRLRWinhomogeneous, bentivegna2016Nbody_InhomogeneityOnExpansion, adamek2013Nbody, mertens2016Nbody_InhomogenousSpacetimes, barrera2020Nbody_gramses, adamek2014Nbody}, and N-body simulations \cite{Barrera2020JCAP:RelativisticNbodyI, Barrera2020JCAP:RelativisticNbodyII, Macpherson2017PhRvD:RelativisticNbody}. While in practice the simulations are still rather limited in extent, any firm progress on the question of the cosmological impact of the inhomogeneous nature of the mass distribution will need such relativistic simulations.


\section{Conclusions}\label{sec:conclusion}
In the past few decades, extensive analyses of cosmic inhomogeneities and their dynamics influencing the behavior of large-scale structures have been undertaken. Most of these backreaction studies utilize either a perturbative approach or a spatially averaging method. We have demonstrated that backreaction exhibits a nested and multiscale character, which refutes the prominent perturbative argument presented by Green \& Wald \cite{green2011framework}. Their claim, asserting the insignificance of backreaction on large scales, has been shown to be untenable. One promising direction for further research is to adopt the perturbative formalism for studying backreaction effects on inhomogeneous structures across various clustering scales.

While scalar averaging approaches, which try to quantify backreaction, have shown numerous promising results in the literature, several mathematical challenges have been overlooked. We have provided a rigorous mathematical treatment on foliations and its connection to gauges in relativity. This analysis has brought to light that standard scalar averaging in a cosmological setting is intrinsically gauge-dependent. In particular, it is significant to consider the averaged particle flow along we average scalar fields. We commented on the notion of averaged particle flow, and noted that they do have a natural relation to cosmic observers, however remain ambiguous to some degree and we leave this question for further research. Furthermore, our mathematical framework for cosmological foliations leads to a clarification of the intricacies from the Gasperini et al. \cite{gasperini2009gauge} averaging approach that has been claimed to be gauge-invariant.

To portray the significance of the cosmic foliation-gauge dependence, we considered a simple toy model of a void. We analyzed the averaged acceleration with respect to the scale factor. We found that different foliation-gauges have a profound effect on the nature of the inferred acceleration within the void model.

To solve this issue for standard cosmological simulations, we built upon the notion of cosmic observers being able to measure the CMB dipole. Transforming away the dipole means that all cosmic observers can find a certain frame, namely a specific kind of cosmic foliation which we call generalized proper time foliations. We show that spatially induced averaging along the averaged particle flow for these generalized proper time foliations is gauge-invariant.

The full implementation on specific cosmological configurations is left to a subsequent paper. This will involve derivations and computations of backreaction effects for cosmic observers that can agree upon this collection of generalized proper time foliations.


\acknowledgments
We thank Syksy Räsänen for a critical appraisal of our manuscript, and thank Robbert W. Scholtens for enjoyable discussions. DV also gratefully acknowledges suitable comments from Sofie Marie Koksbang, and useful discussions with Jaime Pedregal Pastor.


\bibliographystyle{JHEP}
\bibliography{ref.bib}

\end{document}